\documentclass[iop, numberedappendix, appendixfloats]{emulateapj}
\usepackage{graphicx,natbib,amsmath,multirow,hyperref,booktabs}

\shorttitle{Calibrating 100 Years of Polar Faculae Measurements}
\shortauthors{Mu\~noz-Jaramillo, Sheeley, Zhang \& DeLuca}

\begin{document}

\title{Calibrating 100 Years of Polar Faculae Measurements: Implications for the Evolution of the Heliospheric Magnetic Field}

\author{Andr\'es Mu\~noz-Jaramillo\altaffilmark{1, 2, 3, *}, Neil R. Sheeley, Jr.\altaffilmark{4}, Jie Zhang\altaffilmark{5}, and Edward E. DeLuca\altaffilmark{1}}
\affil{$^1$ Harvard-Smithsonian Center for Astrophysics, Cambridge, MA 02138, USA; amunoz@cfa.harvard.edu}
\affil{$^2$ University Corporation for Atmospheric Research, Boulder, CO 80307, USA}
\affil{$^3$ Department of Physics \& Astronomy, University of Utah, Salt Lake City, UT 84112, USA}
\affil{$^4$ Space Science Division, Naval Research Laboratory, Washington, DC 20375-5352, USA}
\affil{$^5$ School of Physics, Astronomy and Computational Sciences, George Mason University, Fairfax, VA 22030, USA}
\affil{$^*$ correspondence should be sent to: \href{mailto:amunoz@cfa.harvard.edu}{amunoz@cfa.harvard.edu}}

\begin{abstract}
Although the Sun's polar magnetic fields are thought to provide important clues for understanding the 11-year sunspot cycle, including the observed variations of its amplitude and period, the current database of high-quality polar-field measurements spans relatively few sunspot cycles.  In this paper we address this deficiency by consolidating Mount Wilson Observatory polar faculae data from four data reduction campaigns, validating it through a comparison with facular data counted automatically from MDI intensitygrams, and calibrating it against polar field measurements taken by the Wilcox Solar Observatory and average polar field and total polar flux calculated using MDI line-of-sight magnetograms.  Our results show that the consolidated polar facular measurements are in excellent agreement with both polar field and polar flux estimates, making them an ideal proxy to study the evolution of the polar magnetic field.  Additionally, we combine this database with sunspot area measurements to study the role of the polar magnetic flux in the evolution of the heliospheric magnetic field (HMF).  We find that there is a strong correlation between HMF and polar flux at solar minimum and that, taken together, polar flux and sunspot area are better at explaining the evolution of the HMF during the last century than sunspot area alone.
\end{abstract}

\keywords{Sun: activity -- Sun: faculae, plages -- Sun: surface magnetism -- Sun: dynamo -- Sun: solar-terrestrial relations}

\section{Introduction}

Solar faculae are bright features on the surface of the Sun associated with accumulations of magnetic flux inside inter-granular lanes (Hale 1922\nocite{hale-1922}).   They are believed to be the consequence of a depression in the optical surface of the Sun caused by the magnetic field, which allows the observer to see the warmer (and thus brighter) walls of the granular upflows (Spruit 1976\nocite{spruit76}, 1977\nocite{spruit77}; Keller et al.\ 2004\nocite{keller-etal04}), and makes them easier to spot near the solar limb.  As a result of this physical relationship, it is not surprising that faculae can be used to track magnetic flux and follow the evolution of surface magnetic fields.  In fact, the numbers of polar facuale, measured on white light photographs taken by the Mount Wilson Observatory (MWO), have been found to be modulated by the solar cycle (Sheeley 1964, 1966\nocite{sheeley64,sheeley66}).  In addition, there is a strong correlation between the polar facular count and the line-of-sight (LOS) polar magnetic field observed by the Wilcox Solar Observatory (WSO; Sheeley 1991, 2008\nocite{sheeley91,sheeley08}).
This correlation, and the ability to see them clearly at the poles, makes faculae as valuable for studying the poloidal aspect of the solar cycle as sunspots are for studying the toroidal part.  The goal of this study is to standardize, validate, and magnetically calibrate the measurements of polar faculae observed at MWO during the past hundred years.

\begin{figure*}
\centering
  \includegraphics[scale=0.4]{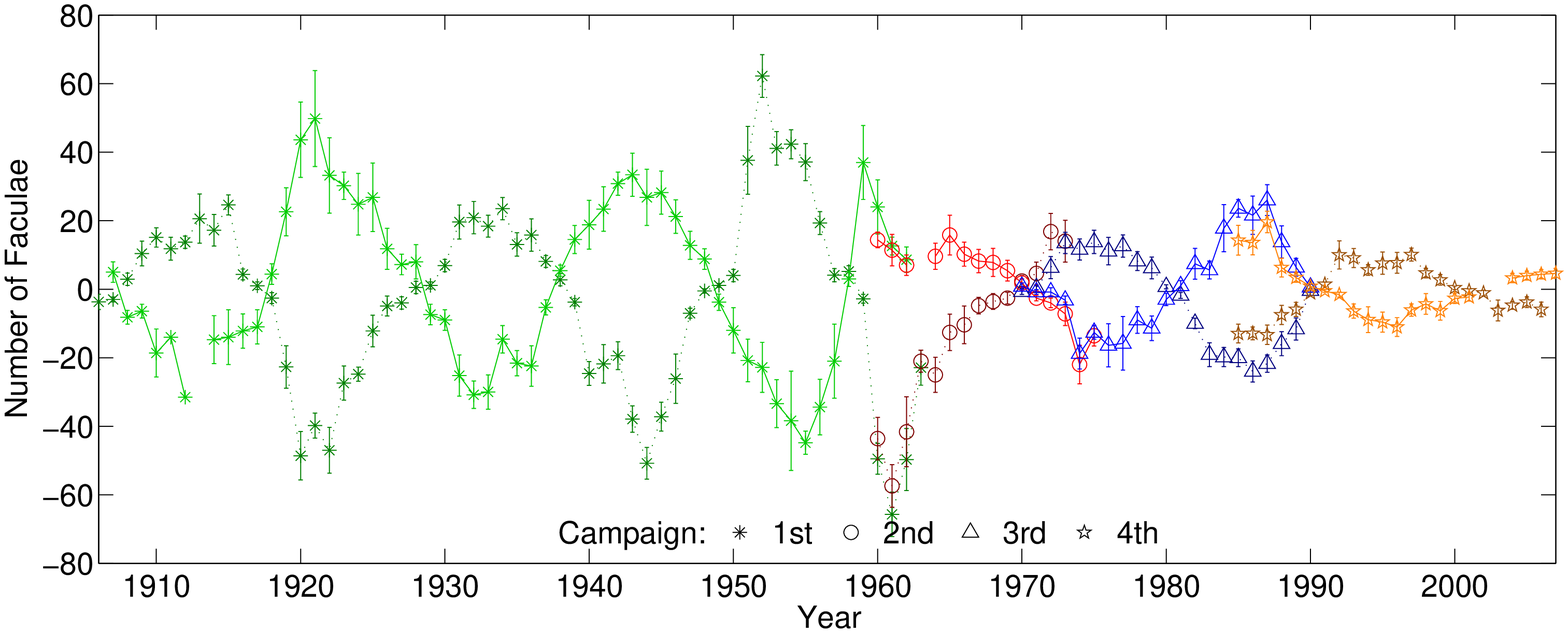}
  \caption{Raw MWO Facular measurements.  Each campaign is marked with a different color and marker: green asterisk (1st), red circle (2nd), blue triangle (3rd), and orange star (4th). Measurements for the north (south) pole are shown using a dark dashed (light solid) line.  After each minima the sign is reversed to match the polarity of each magnetic cycle.}\label{Fig_MWO_Raw}
\end{figure*}

\begin{figure*}
\centering
\begin{tabular}{c}
\begin{tabular}{cc}
  \includegraphics[scale=0.4]{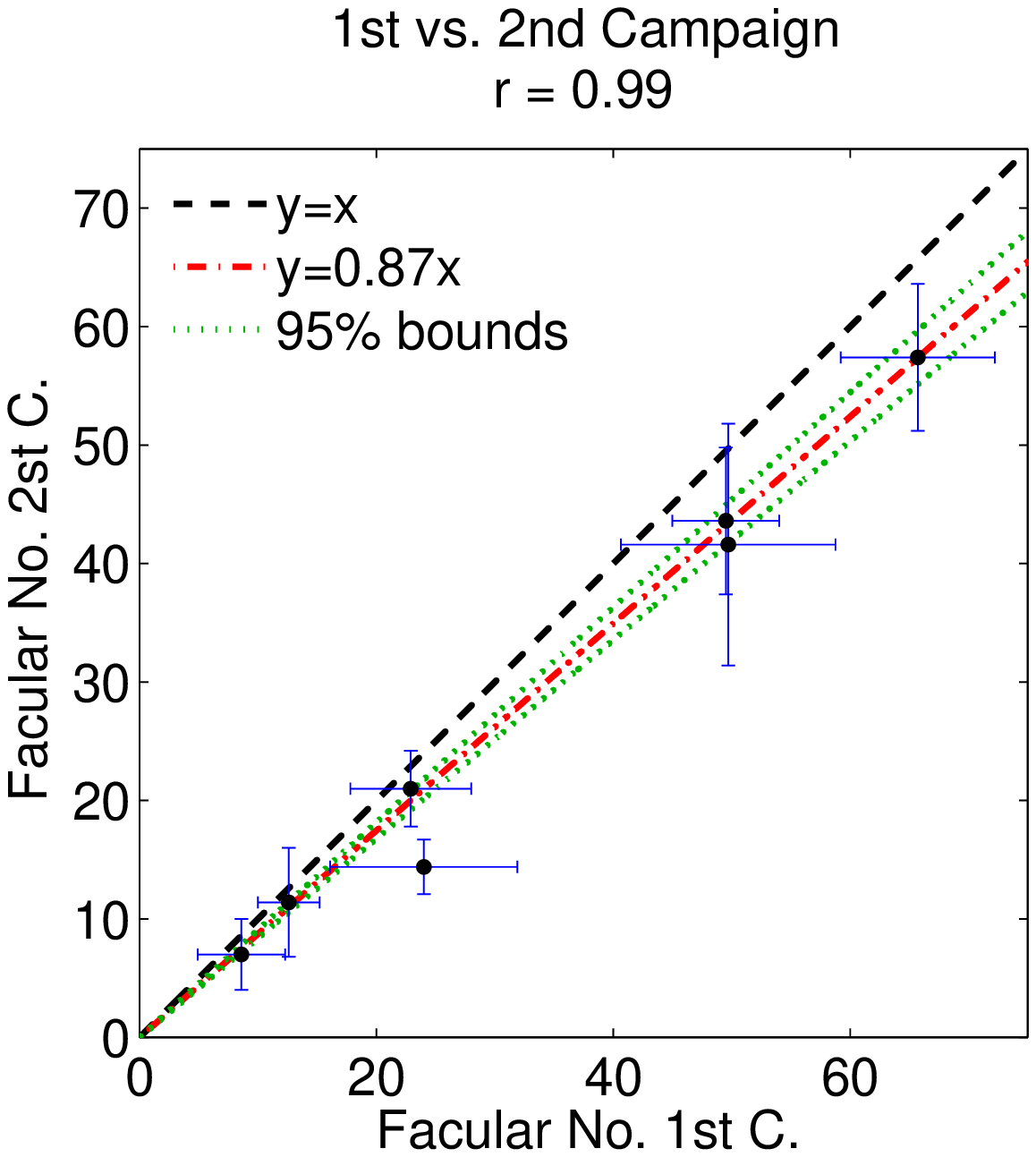} & \includegraphics[scale=0.4]{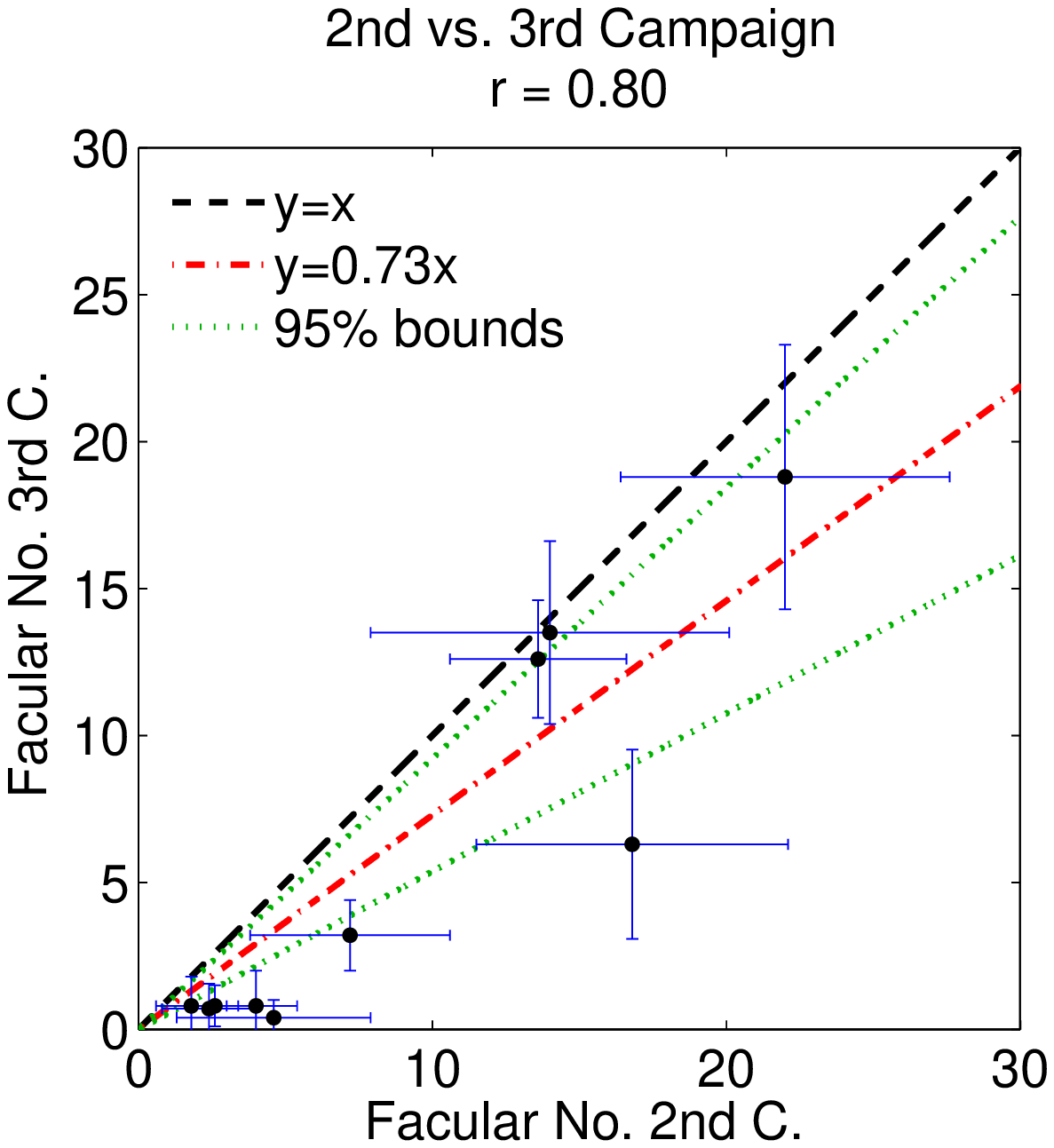} \\
                    (a)                              &                   (b)
\end{tabular}\\
  \includegraphics[scale=0.4]{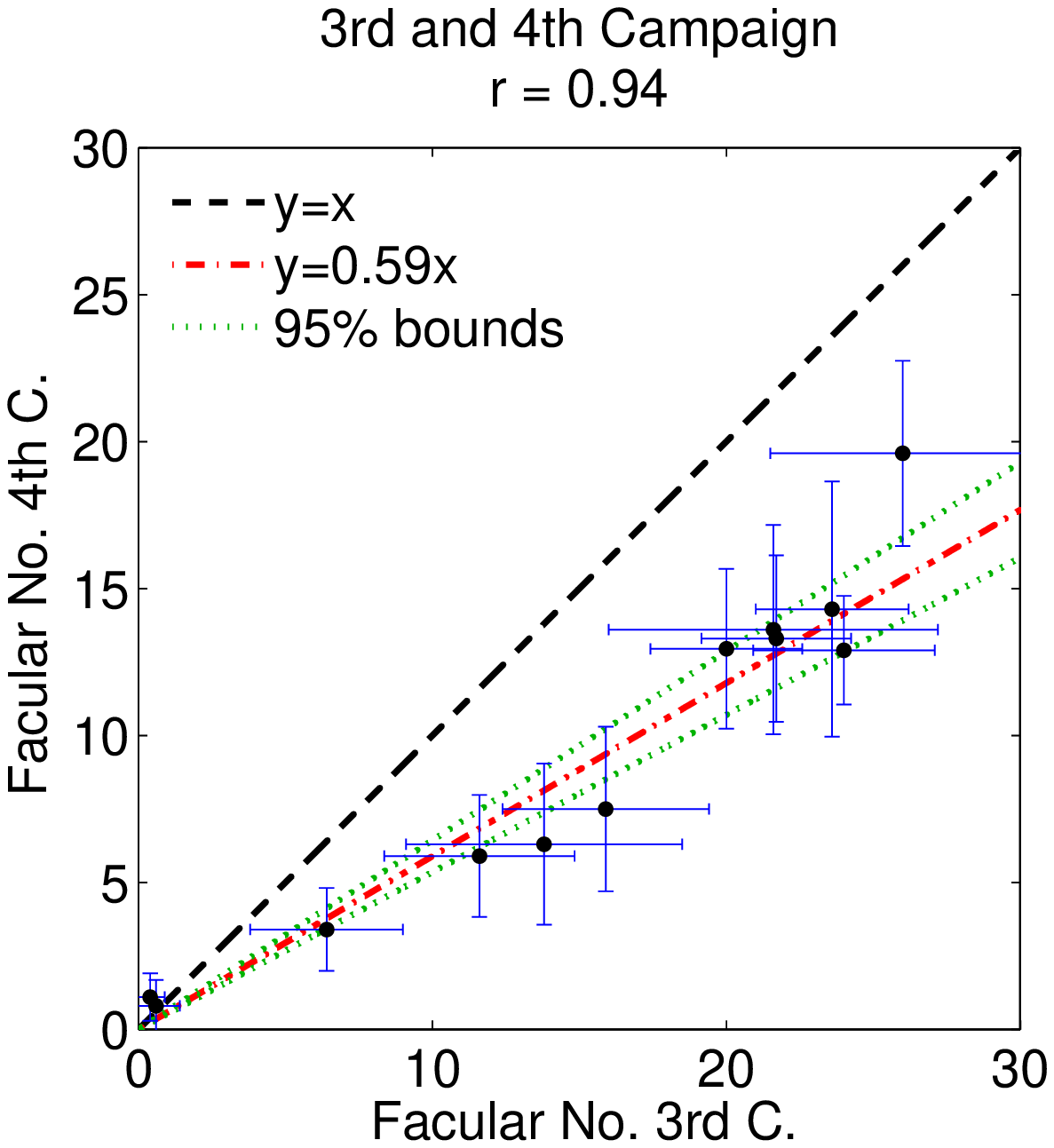}\\
                        (c)
\end{tabular}
\caption{(a) Scatter plot comparing MWO facular measurements across data reduction campaigns: (a) 1st vs.\ 2nd, (b) 2nd vs.\ 3rd, and (c) 3rd vs.\ 4th.  Each scatter plot is fitted with a line passing through the origin (dot-dashed red).  The 95\% confidence interval is bounded by dotted green lines and a line of slope one is plotted in dashed black for reference.  The factors used to calibrate the data across different campaigns are: $0.87\pm0.03$ between the 1st and 2nd, $0.73\pm0.19$ between 2nd and 3rd, and $0.59\pm0.05$ between 3rd and 4th. They correspond to the slopes and the 95\% confidence interval of their respective linear fits.}\label{Fig_MWO_Cal}
\end{figure*}

\section{Cross-Calibration of MWO Polar Faculae Measurements}\label{Sec_MWO_Cal}

In this work we use MWO polar facular measurements taken in four different data reduction campaigns (1906-1964, Sheeley 1964, 1966; 1960-1975, Sheeley 1976; 1970-1990, Sheeley 1991; and 1985-2007, Sheeley 2008)\nocite{sheeley64,sheeley66,sheeley76,sheeley91,sheeley08}.  Two sets of high-quality white light photographs were selected during each data reduction campaign, each corresponding to the intervals when the solar poles are most visible from Earth (Feb-15 to Mar-15 for the south pole and Aug-15 to Sep-15 for the north pole).  The five best images were selected for each interval based on image quality, photographic contrast, and (whenever possible) that such high-quality images be uniformly spaced in time.   Polar faculae were counted based on eye estimates of bright features whose contrasts were comparable to those of low latitude faculae, but whose sizes were much smaller. To remove possible biases, the images in each deck (either spring or fall) were shuffled before obtaining the facular count, and the dates of observation were recorded afterward.  Then the measurements were reorganized in chronological order and the five measurements for each year were averaged.  In our study we use only the facular measurements for the most visible pole (south or north, respectively for the spring and fall decks).  Note that though facular measurements are a strictly positive quantity, they have been given a sign (which reverses when they reach a minimum) to match the polarity believed (or known) to be present at each pole during each polar cycle.  This allows for easier visualization and for comparison with signed quantities like magnetic field and flux.  It also removes discontinuities in the slope of the faculae/time curve at times that the field goes through zero.  However, this sign is not taken into consideration when comparing different facular measurements.

\begin{figure*}
\centering
  \includegraphics[scale=0.4]{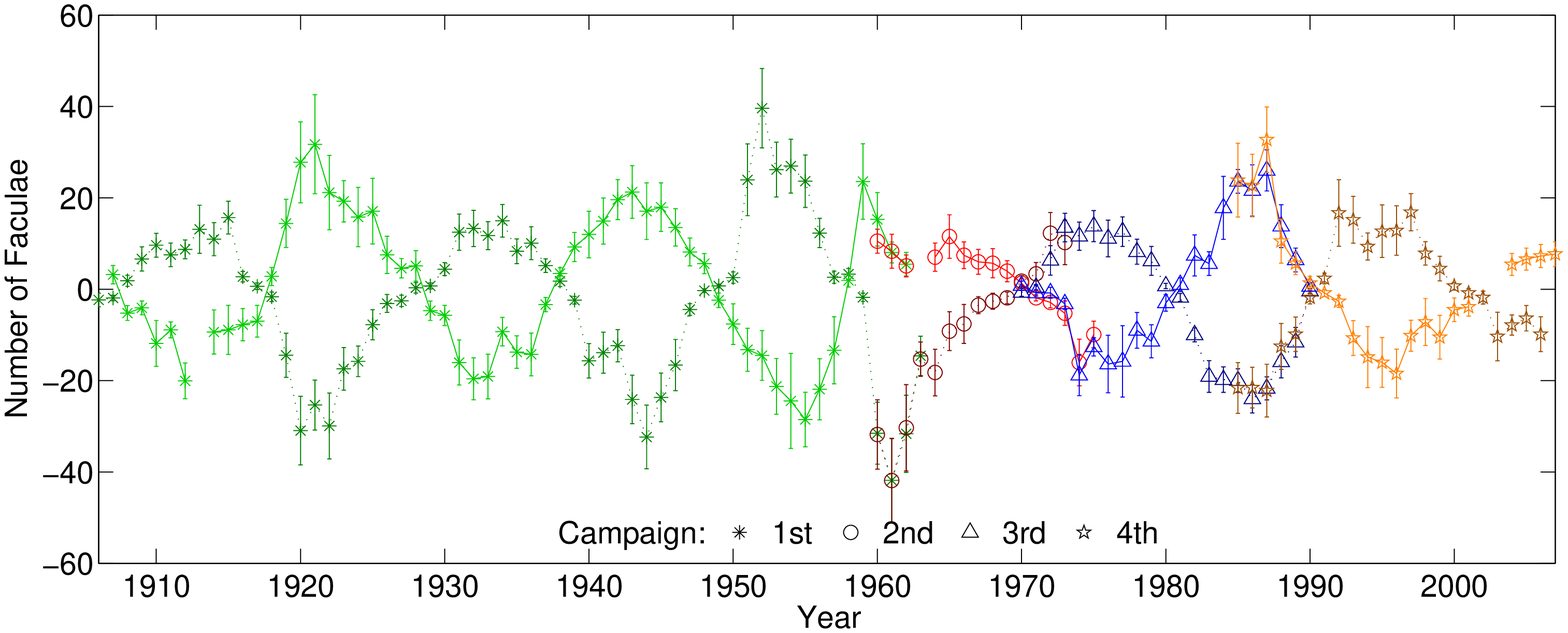}
  \caption{Calibrated MWO Facular measurements.  Each campaign is marked with a different color and marker: green asterisk (1st), red circle (2nd), blue triangle (3rd), and orange star (4th). Measurements for the north (south) pole are shown using a dark dashed (light solid) line.  All campaigns are calibrated to the 3rd campaign.}\label{Fig_MWO_Cal2}
\end{figure*}

As can be seen in Figure \ref{Fig_MWO_Raw}, there is some overlap between campaigns.  This was arranged intentionally to provide a means for intercalibration measurements from consecutive campaigns.  When we performed this calibration, we found a systematic effect in which the measured values became smaller from one campaign to the next.  This discrepancy is relatively small between the 1st and 2nd campaigns and the 2nd and 3rd campaigns, but it is larger between the 3rd and 4th campaigns, as shown in Figure \ref{Fig_MWO_Cal}.  It is difficult to estimate exactly the source of this discrepancy given the subjective nature of the measurements, but the factors may include the use of a different location for making the measurements during the 4th campaign (with a different light box and different viewing conditions), and a stricter criterion for identifying faculae with the passage of time.   Fortunately, the goodness of a linear fit through each overlap interval (0.99 for 1st and 2nd campaigns, 0.80 for the 2nd and 3rd, and 0.94 for the 3rd and 4th) suggests that these discrepancies can be corrected with a multiplicative factor corresponding to the slope of the linear fit.  Furthermore, the fact that the 3rd and 4th observations are well within the interval of WSO observations allows us to verify that this correction is quite adequate in the case of the campaigns with the biggest discrepancy (3rd and 4th; see Figure \ref{Fig_MWO_WSO_MDI}).  In this work we use the 95\% confidence intervals as a measure of the error in the multiplicative factors and reference all data reduction campaigns to the 3rd campaign in order to minimize error propagation.

The adequacy of the multiplicative correction is further confirmed in Figure \ref{Fig_MWO_Cal2}, where we can observe very good agreement across different campaigns.  This leaves us with a standardized polar faculae dataset covering more than a century of observations.  The next step is to validate these results using an automatic algorithm for counting polar faculae on data taken with the Michelson Doppler Imager (MDI; Scherrer et al.\ 1995\nocite{scherrer-etal95}) on the solar and heliospheric observatory (SOHO) spacecraft.  This way we ensure that the MWO facular measurements are robust and not subject to methodological or instrumental errors, and we validate the number of polar faculae counted in the interval 2002-2008 when high-quality photographic film was no longer commercially available for obtaining the MWO images.

\begin{figure}
\centering
  \includegraphics[width=0.47\textwidth]{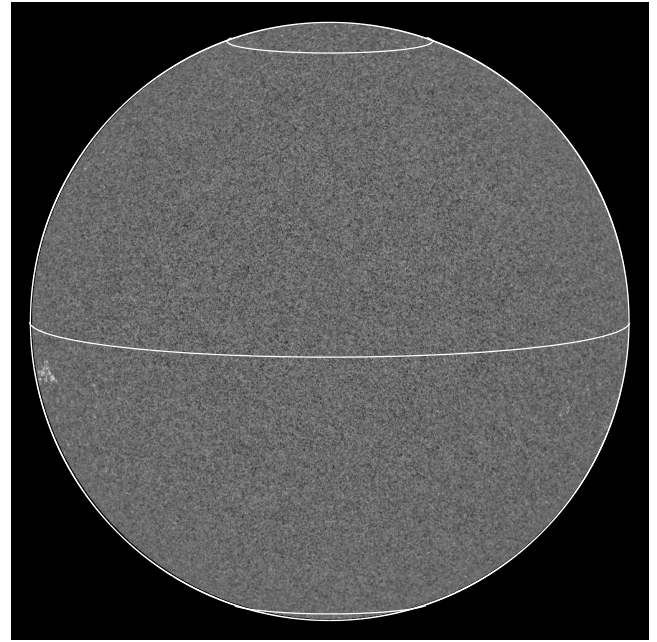}\\
  \caption{MDI 2.0 intensitygram taken on 20-Aug-2007. The circles shown in the image correspond to $70^o$, $0^o$ and $-70^o$ in heliographic latitude.}\label{Fig_MDI_I}
\end{figure}

\section{Validation of Mount Wilson Polar Faculae Measurements using MDI}\label{Sec_MDI_val}

\begin{figure}
\centering
\begin{tabular}{c}
  \includegraphics[width=0.47\textwidth]{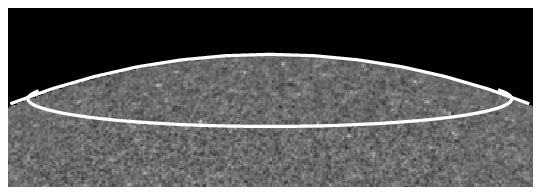}\\
  (a)\\
  \includegraphics[width=0.47\textwidth]{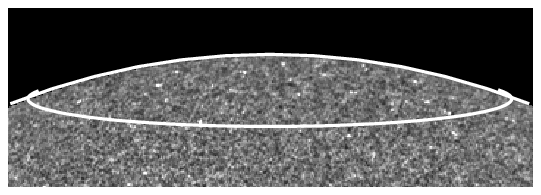}\\
  (b)\\
  \includegraphics[width=0.47\textwidth]{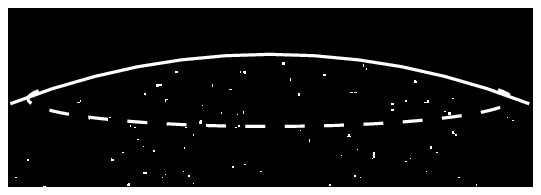}\\
  (c)\\
  \includegraphics[width=0.47\textwidth]{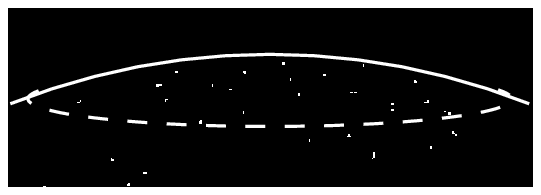}\\
  (d)
\end{tabular}
\caption{Steps in the automatic detection of polar faculae.  First the MDI data (a) are scaled using a gamma function (b).  Then a mask is built based on a threshold (c) and filtered for individual pixels (d).  Polar faculae are counted automatically above $70^o$ for the north pole and below $-70^o$ for the south pole.  The images show a $70^o$ latitude line for illustration purposes.}\label{Fig_Fac_P}
\end{figure}

In this work we use continuum intensitygrams taken by MDI going from 19-May-1996 to 26-Dec-2010.   Each intensitygram image is 1024x1024 pixels from which limb-darkening has been removed; see Figure~\ref{Fig_MDI_I}).
We use one image per day (when available), counting the number of faculae above (below) 70$^o$N (70$^o$S) using an automatic detection code based on that of Zhang, Wang and Liu (2010\nocite{zhang-wang-liu10}). For our work the regions poleward of 70$^o$ are defined as the ``pole''.  The algorithm we use has the following steps:
\begin{enumerate}
  \item Perform gamma scaling of the image to enhance contrast using the expression $f(x)= k_0 x^{\gamma}$, where $\gamma=15$ is the enhancement exponent and $k_0=100$ is a normalization constant (See Figs.~\ref{Fig_Fac_P}-a \& \ref{Fig_Fac_P}-b).  These values are chosen considering that the average of each MDI intensitygram is close to 1.
  \item Mask pixels above a certain intensity threshold ($160.0$) which are also located at each pole (See Fig.~\ref{Fig_Fac_P}-c).  Due to excessive limb brightening, the outer three radial pixels on the solar disk (corresponding to an angular size of 6") are ignored.
  \item Remove single facular pixels in order to distinguish facular regions from small and bright intergranular regions (See Fig.~\ref{Fig_Fac_P}-d)
  \item Count each isolated facula, fully automatically, independently of the amount of pixels it contains.
\end{enumerate}

After removing overexposed, overcorrected and incomplete images we obtain a daily data series of facular count (see top panels of Figures \ref{Fig_Fac_D}-a \& b), to which we apply a month-long running mean (see bottom panels of Figures \ref{Fig_Fac_D}-a \& b).  This way, we sample a time interval of approximately the same length as the one sampled during the MWO data reduction campaigns (Sheeley 1964, 1966, 1976, \& 2008).  Finally, in order to validate the Mount Wilson facular count, we select averaged facular counts corresponding to 4-Mar (4-Sep) of each year for the south (north) pole (see Figure \ref{Fig_Fac_D}-c).

\begin{figure*}
\centering
\begin{tabular}{cc}
  \includegraphics[scale=0.42]{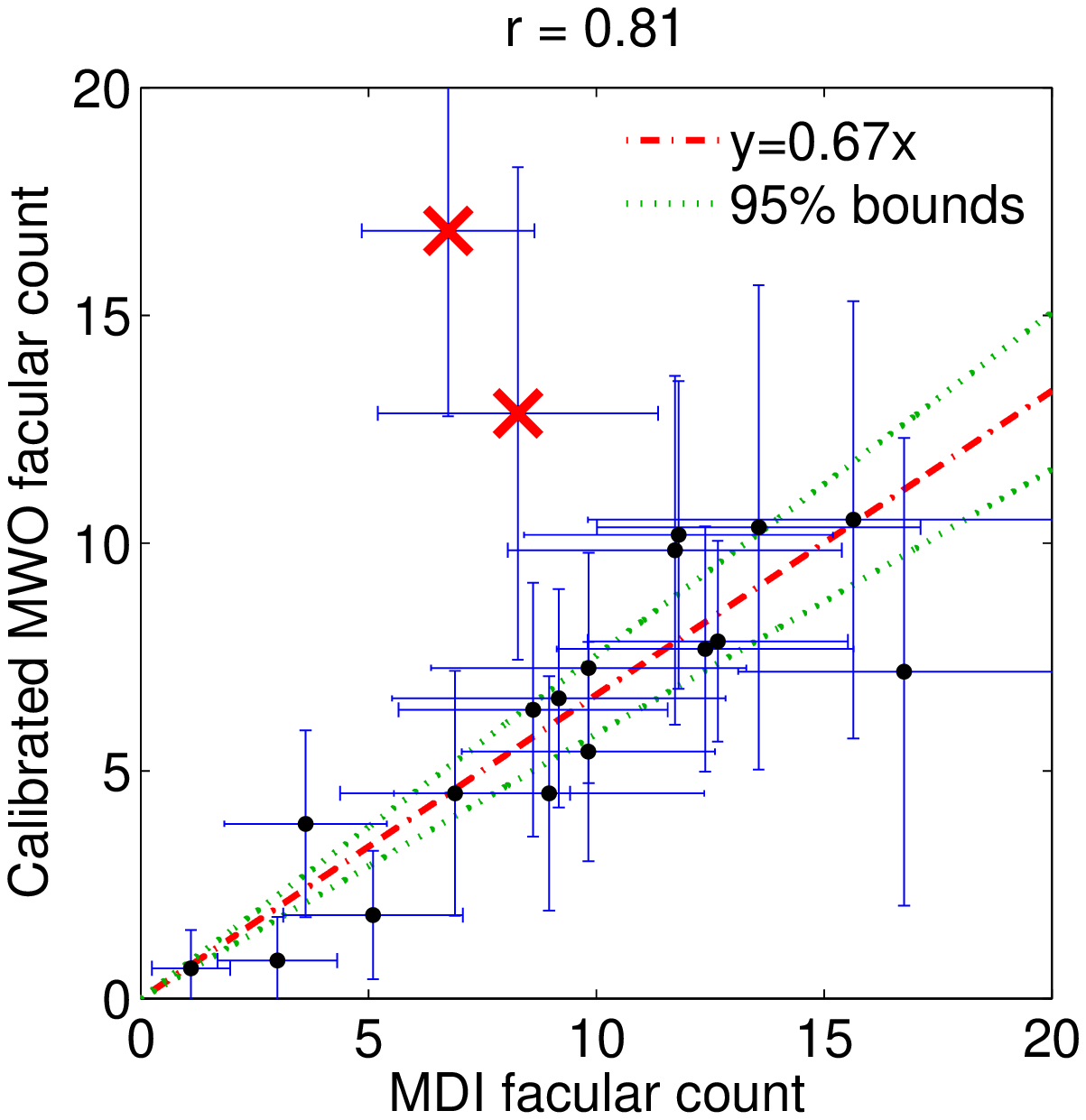} & \includegraphics[scale=0.42]{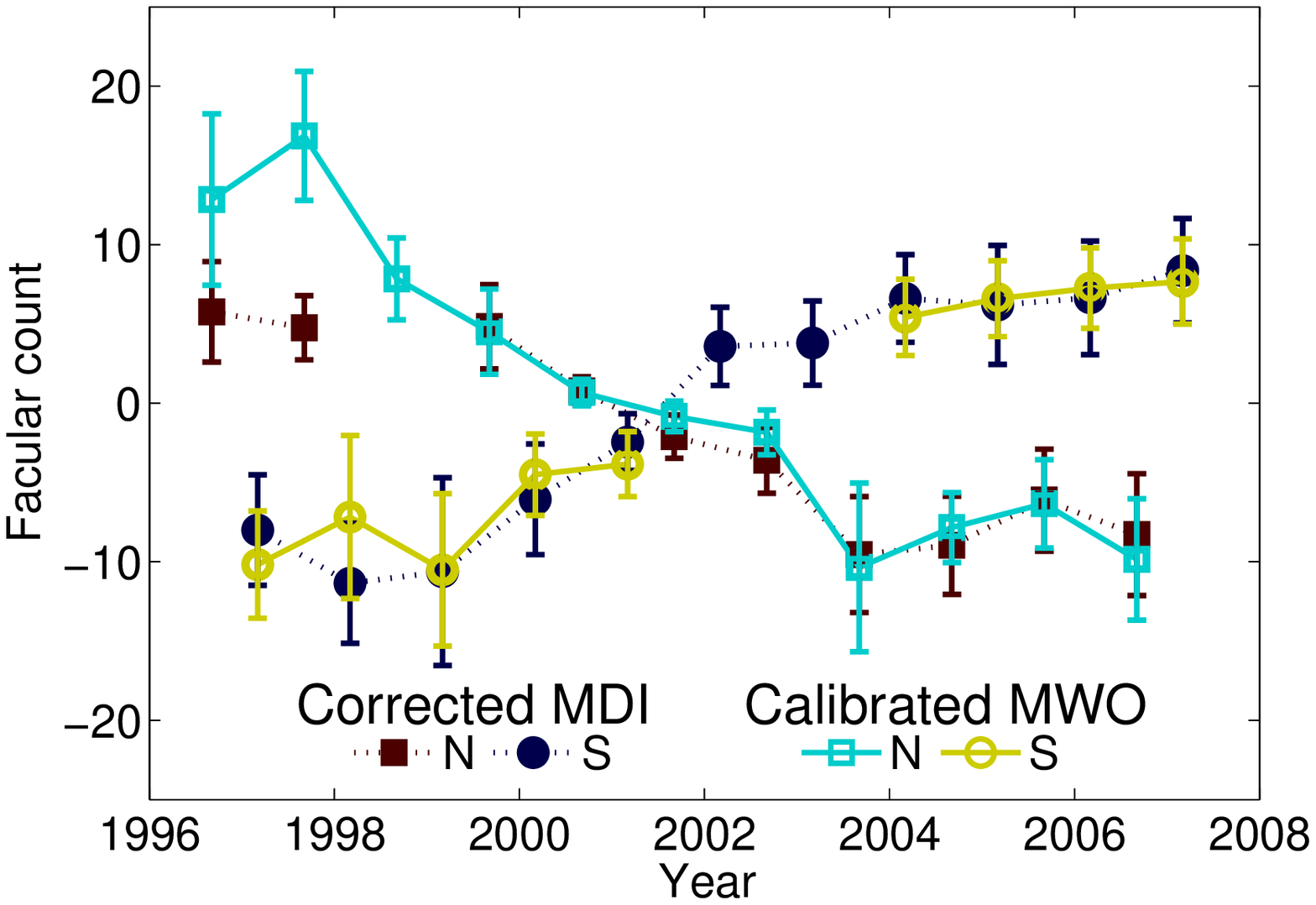}\\
                    (a)                              &                   (b)
\end{tabular}
\caption{(a) Scatter plot comparing MDI and Mount Wilson facular measurements. The data (minus outliers) are fit very well ($r=0.81$) by a line with a slope of $1.06$ which passes trough the origin.  (b)  Overplot of the calibrated Mount Wilson and the MDI corrected facular measurements.  MDI measurements are corrected by a factor of $0.67\pm0.08$ in the figure corresponding to the slope and the 95\% confidence interval of the linear fit.  The outliers are clearly observed as the MDI facular measurements of 1996 and 1997.}\label{Fig_Fac_MWO_MDI}
\end{figure*}

As can be seen in Figure \ref{Fig_Fac_MWO_MDI}-a, MWO and MDI facular measurements are in good agreement with the exception of two points (marked with an x in Fig.~\ref{Fig_Fac_MWO_MDI}-a); this is likely due to different resulting contrasts in the MDI calibration before and after contact was lost with SOHO.  Nevertheless, it is reassuring that results obtained by such different methods differ only by a multiplicative factor (see Fig.~\ref{Fig_Fac_MWO_MDI}-a).  Furthermore, not only do the measurements agree very well in their actual values, but their relative errors are also essentially the same.  This means that even though the Mount Wilson measurements only use five points per averaging month (as opposed to a daily measurement in the MDI facular measurement), they are a representative sample of such an interval and are sufficient to capture polar facular variability in the measuring period.  Taken together, these results strongly support the validity and relevance of the combined Mount Wilson dataset. The next step is to use the MWO data to estimate the long term evolution of the polar magnetic properties.

\section{Comparison Between MWO Facular Measurements, WSO Polar Field Measurements, and MDI LOS Magnetograms}

Given that ultimately our objective is to gain insight into the long term evolution of the solar magnetic cycle, it is necessary to evaluate the adequacy of polar facular data as a proxy for the polar magnetic field strength and for the signed polar magnetic flux.  Here we use polar field measurements taken by the WSO, which have already been correlated favorably with the MWO dataset (Sheeley 2008\nocite{sheeley08}). These data have been taken over a period of time that overlaps the four polar faculae data reduction campaigns.  We apply a correction factor  (a 15\% increase) to the WSO measurements taken during the first two years of observations (1976-1977), in order to account for scattered light caused by dusty mirrors and lenses; Svalgaard \& Hoeksema personal communication.  We also use MDI magnetograms which have directly associated facular measurements and are available at a higher cadence (15 per day); we process MDI magnetograms along similar guidelines as MDI intensitygrams.

\subsection{Polar Faculae as a Proxy for Polar Magnetic Field Strength}

\begin{figure*}
\centering
\begin{tabular}{c}
  \includegraphics[scale=0.46]{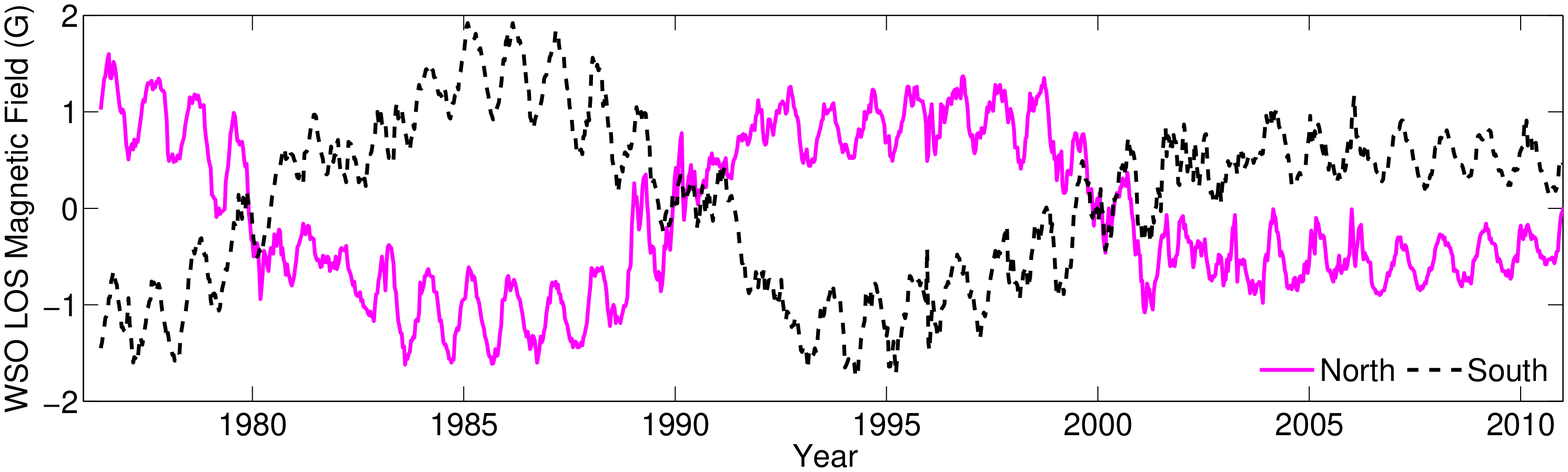}\\
  (a)\\
  \includegraphics[scale=0.46]{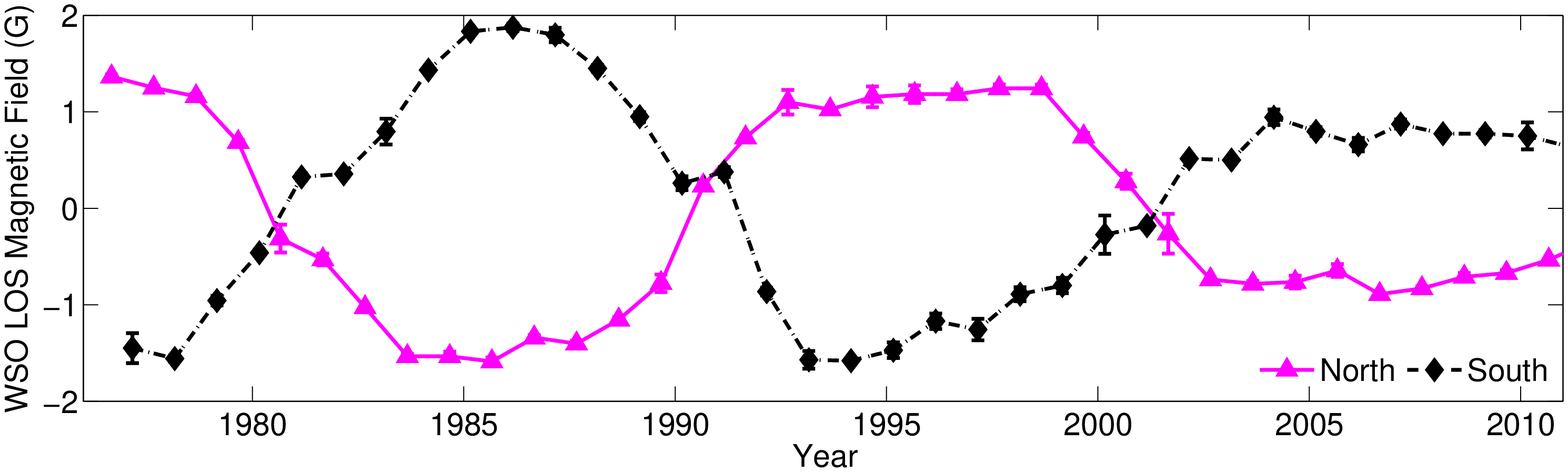}\\
  (b)
\end{tabular}
\caption{(a) Wilcox solar observatory measurements of the polar field. A solid magenta (black dashed) line corresponds to the north (south) pole magnetic field. (b) Selected polar field averages. Points corresponding to the 4-Mar (4-Sep) of each year are used for the south (north) pole.  These dates are chosen for consistency and comparison with other datasets.}\label{Fig_WSO}
\end{figure*}

Because WSO magnetic field measurements span 35 years (beginning at 1976; see Figure \ref{Fig_WSO}-a), they are the ideal reference dataset.  The polar field strength for both the MWO polar faculae measurements and the MDI average polar field measurements are calibrated to their WSO counterparts. To develop a consistent set of measurements, we use the average of measurements taken between the 15-Aug and the 15-Sep for the north pole and between 15-Feb and 15-Mar for the south pole (see Figure \ref{Fig_WSO}-b).

\begin{figure}
\centering
  \includegraphics[width=0.47\textwidth]{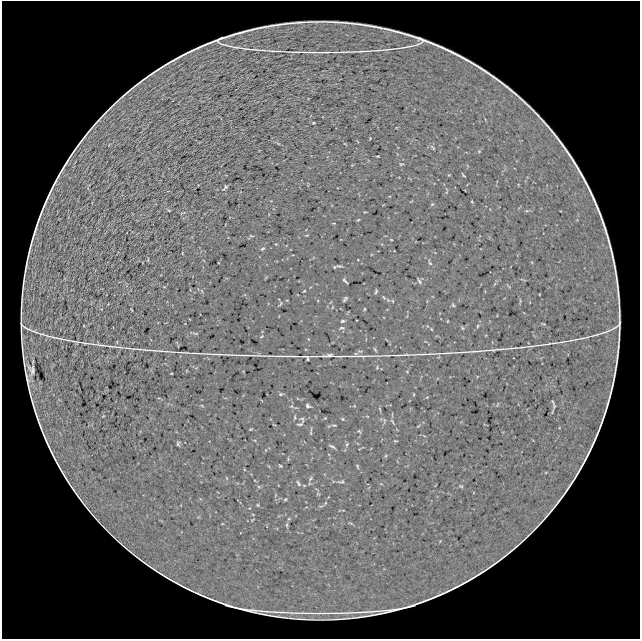}
  \caption{MDI 1.8 line-of-sight magnetogram taken on 20-Aug-2007. The circles shown in the image correspond to $70^o$, $0^o$ and $-70^o$. Colors are saturated in order to enhance visibility of polar patches of magnetic field.}\label{Fig_MDI_LOS_Raw}
\end{figure}

\begin{figure*}
\centering
\begin{tabular}{c}
  \includegraphics[scale=0.44]{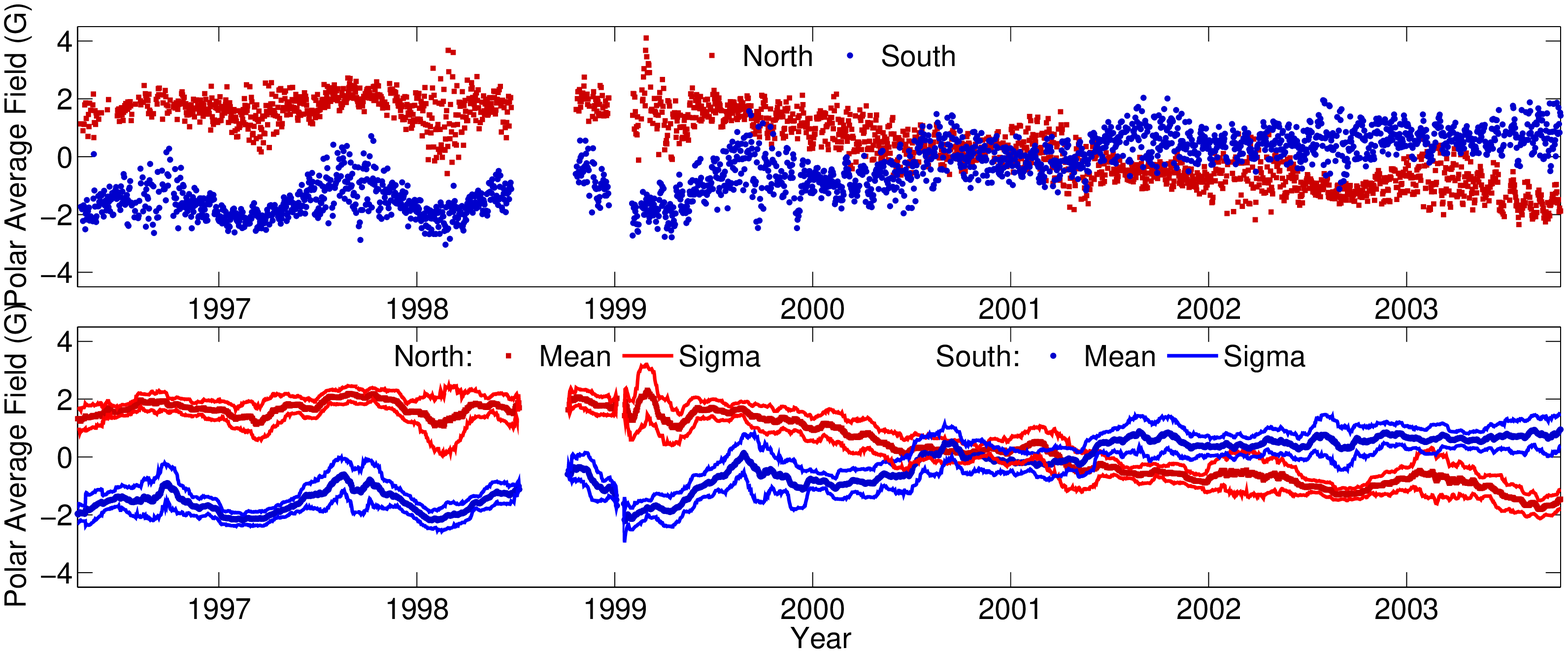}\\
  (a)\\
  \includegraphics[scale=0.44]{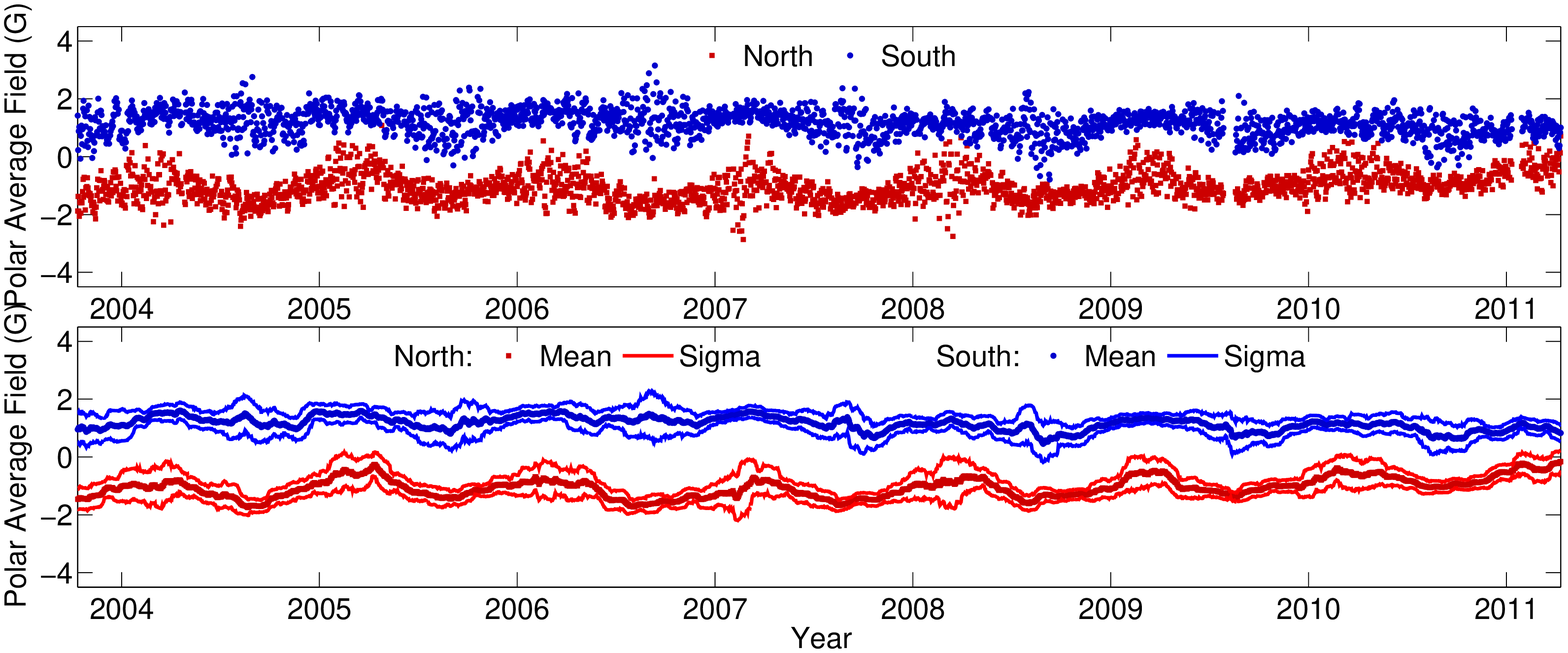}\\
  (b)\\
  \includegraphics[scale=0.44]{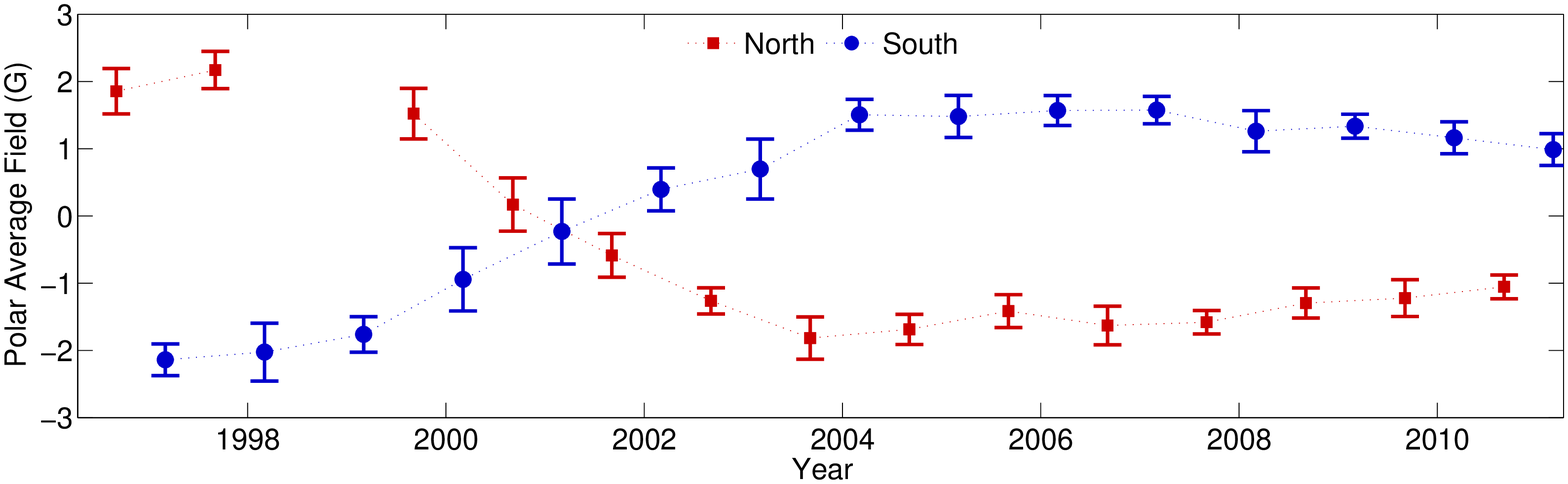}\\
  (c)
\end{tabular}
\caption{MDI average LOS magnetic above field above (below) $70^o$ ($-70^o$) for the 1996-2004 (a) and 2004-2011 (b) intervals.  The top panel of each figure shows the daily average field; the bottom panel shows average field after being applied a month-long running average.  Red (blue) corresponds to the north (south) pole and the thin lines correspond to one standard deviation. (c) Selected values of the average LOS magnetic field for the 1996-2011 interval.  Points corresponding to the 4-Mar (4-Sep) of each year are used for the south (north) pole.  These dates correspond to the point in Earth's orbit with the best observations of each particular pole.}\label{Fig_MDI_LOS_Pros}
\end{figure*}

We calculate the average polar field strength using MDI line-of-sight (LOS) magnetograms (Figure \ref{Fig_MDI_LOS_Raw} shows the MDI LOS magnetogram associated with the intensitygram shown in Figure \ref{Fig_MDI_I}). For 371 days in our time period no data was available (most of them during the so-called ``MDI vacation").  We select one magnetogram per day and calculate the average LOS magnetic field polarward of $70^o$ (see top panels of Figures \ref{Fig_MDI_LOS_Pros}-a \& b), perform a month-long running average (see bottom panels of Figures \ref{Fig_MDI_LOS_Pros}-a \& b), and select the values of the average polar LOS magnetic field corresponding to 4-Mar (4-Sep) of each year for the south (north) pole (see Figure \ref{Fig_MDI_LOS_Pros}-c).  This standardizes the measurements to be consistent with other data used in this work.

\begin{figure*}
\centering
\begin{tabular}{c}
  \includegraphics[scale=0.44]{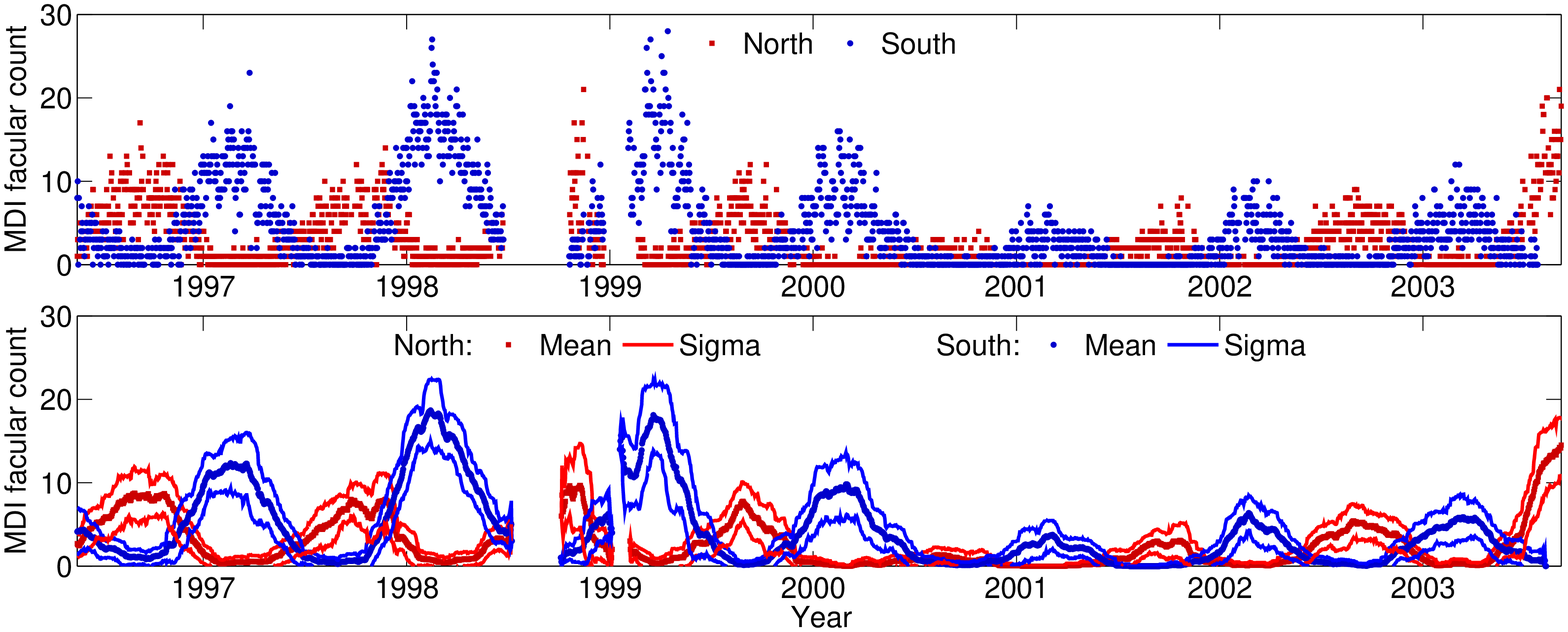}\\
  (a)\\
  \includegraphics[scale=0.44]{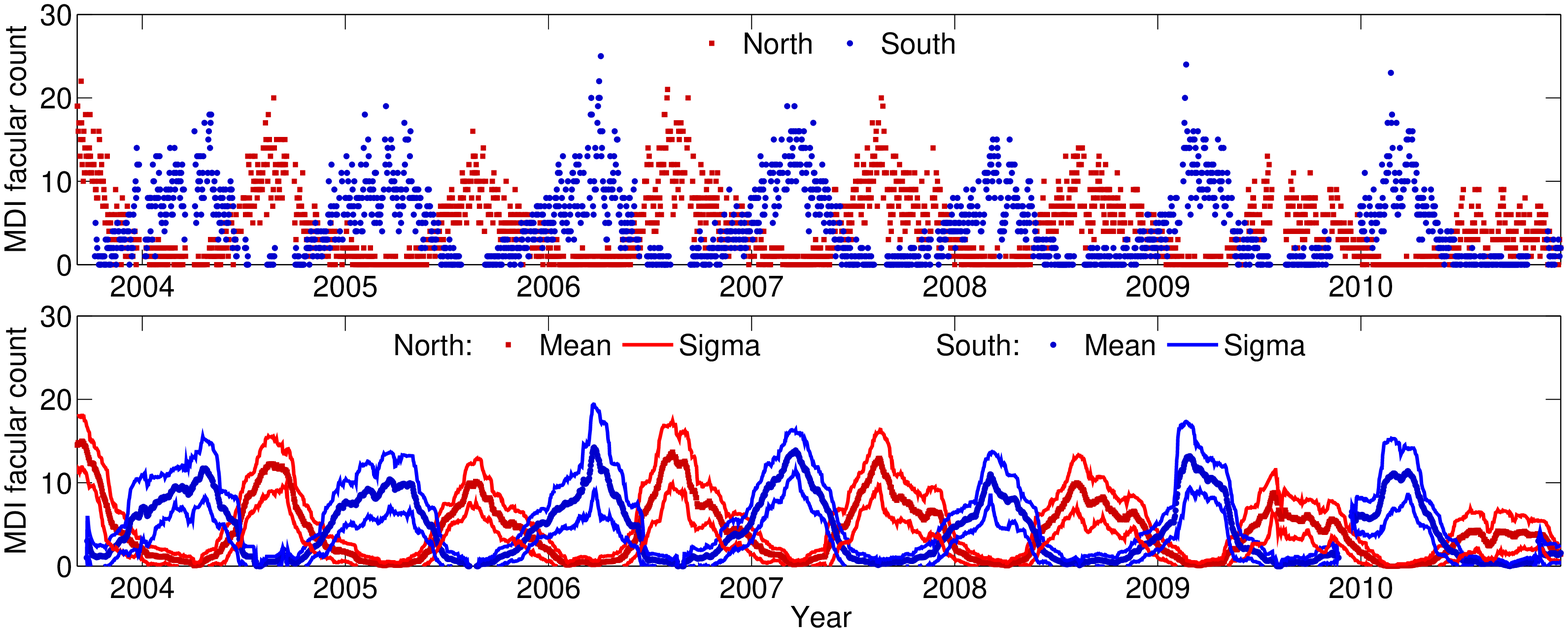}\\
  (b)\\
  \includegraphics[scale=0.44]{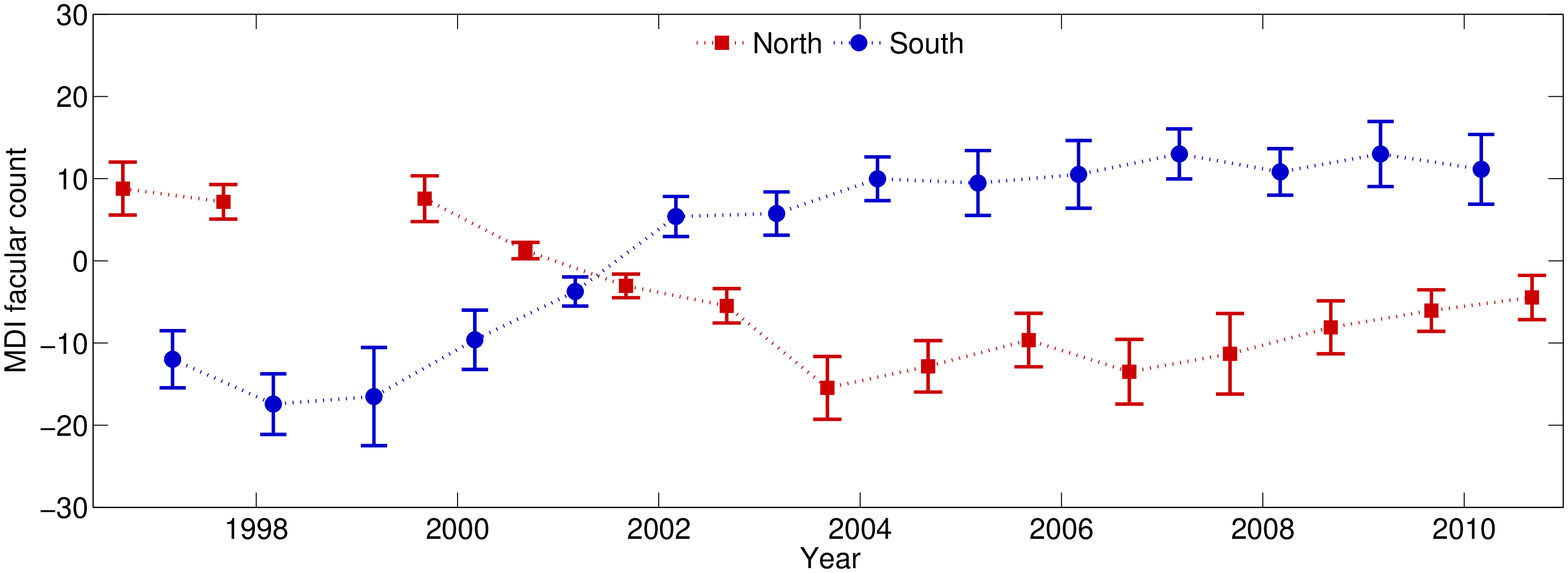}\\
  (c)
\end{tabular}
\caption{MDI facular count for the 1996-2004 (a) and 2004-2011 (b) intervals.  The top panel of each figure shows the daily facular count; the bottom panel shows facular count after being applied a month-long running average.  Red (blue) corresponds to the north (south) pole and the thin lines correspond to one standard deviation.  (c) Selected facular averages for the 1996-2011 interval.  Points corresponding to the 4-Mar (4-Sep) of each year are used for the south (north) pole.  These dates correspond to the point in Earth's orbit with the best observations of each particular pole.
}\label{Fig_Fac_D}
\end{figure*}

\begin{figure*}
\centering
\begin{tabular}{c}
\begin{tabular}{cc}
  \includegraphics[scale=0.4]{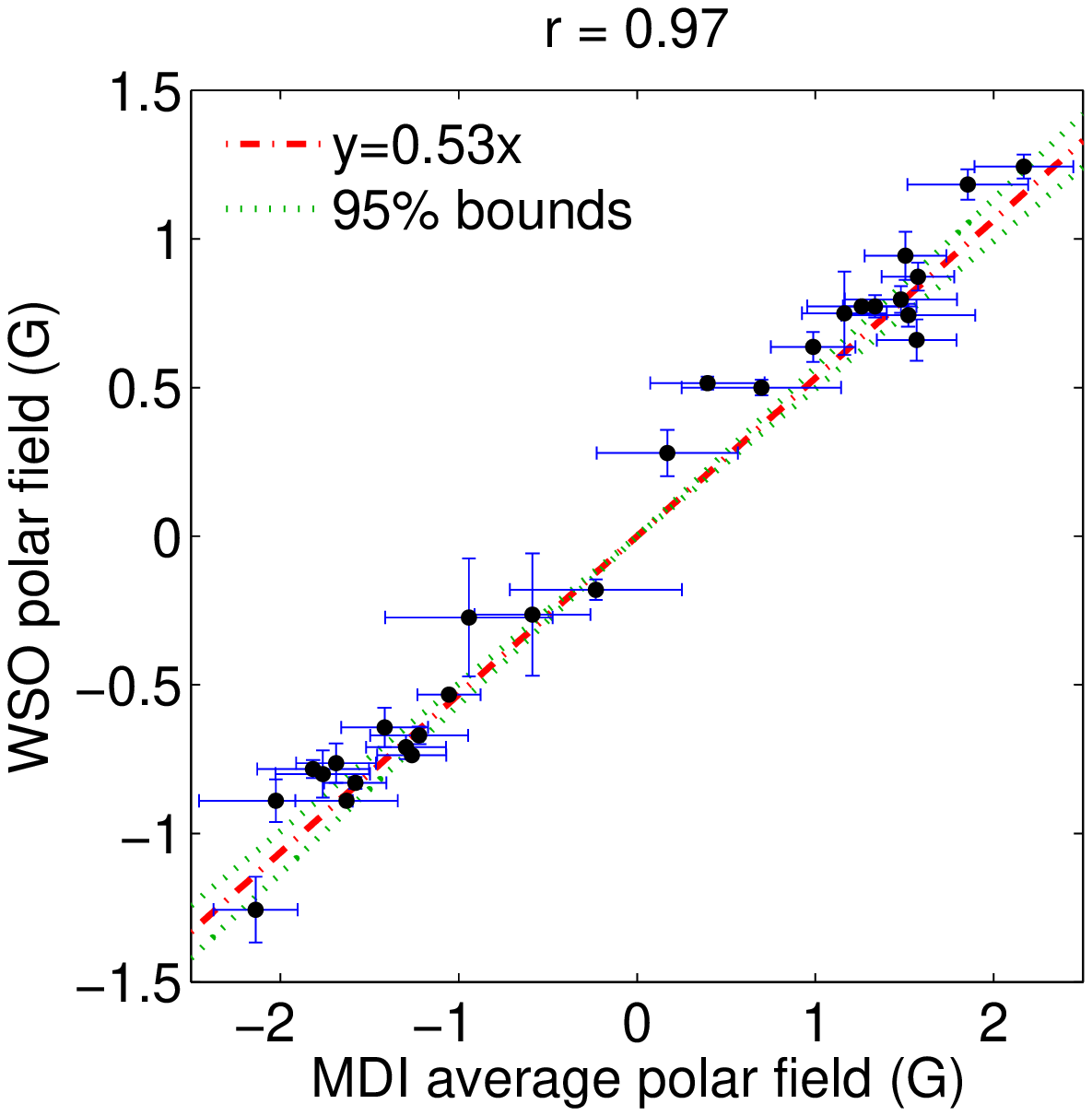} & \includegraphics[scale=0.4]{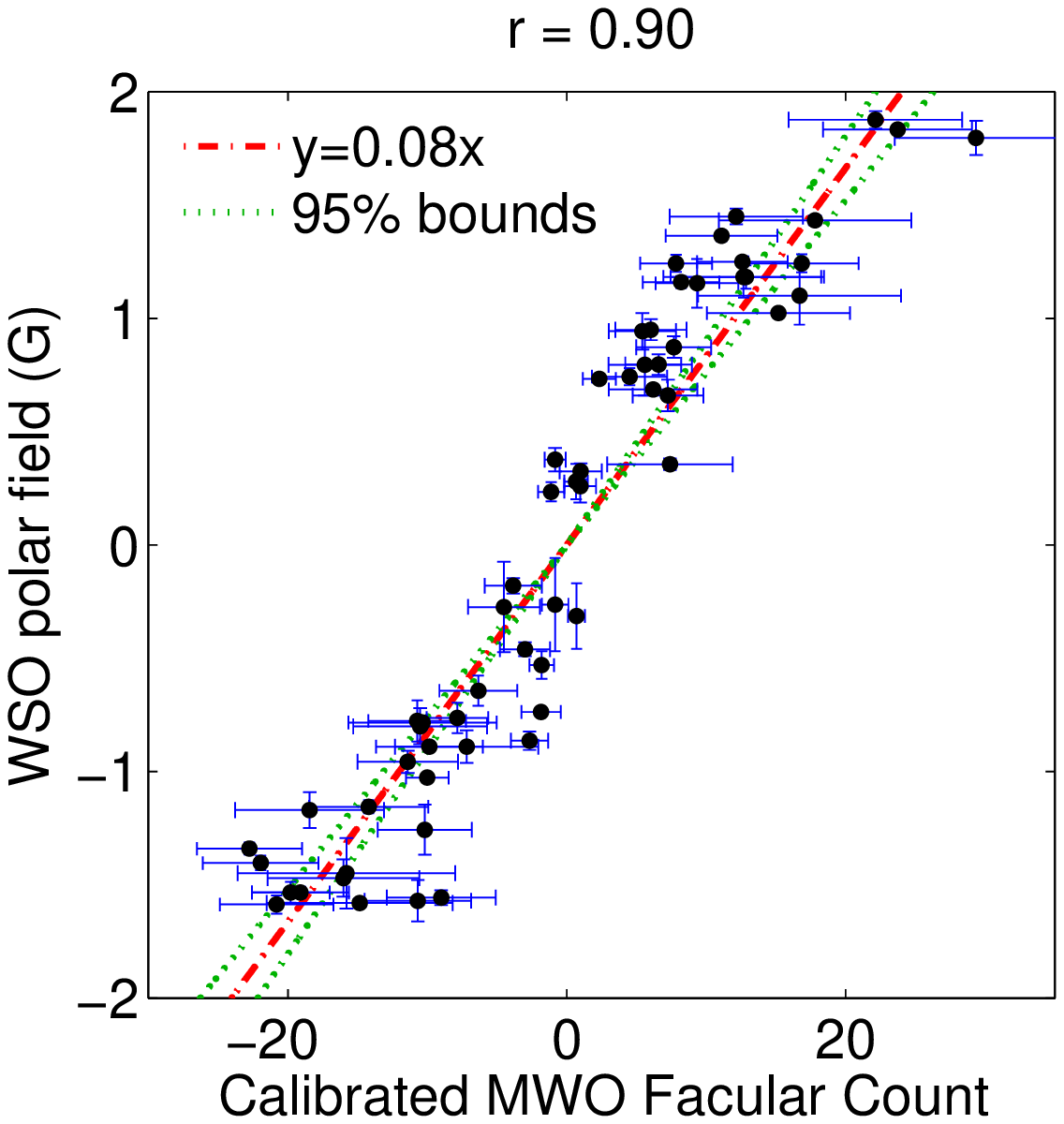} \\
                    (a)                              &                   (b)
\end{tabular}\\
  \includegraphics[scale=0.35]{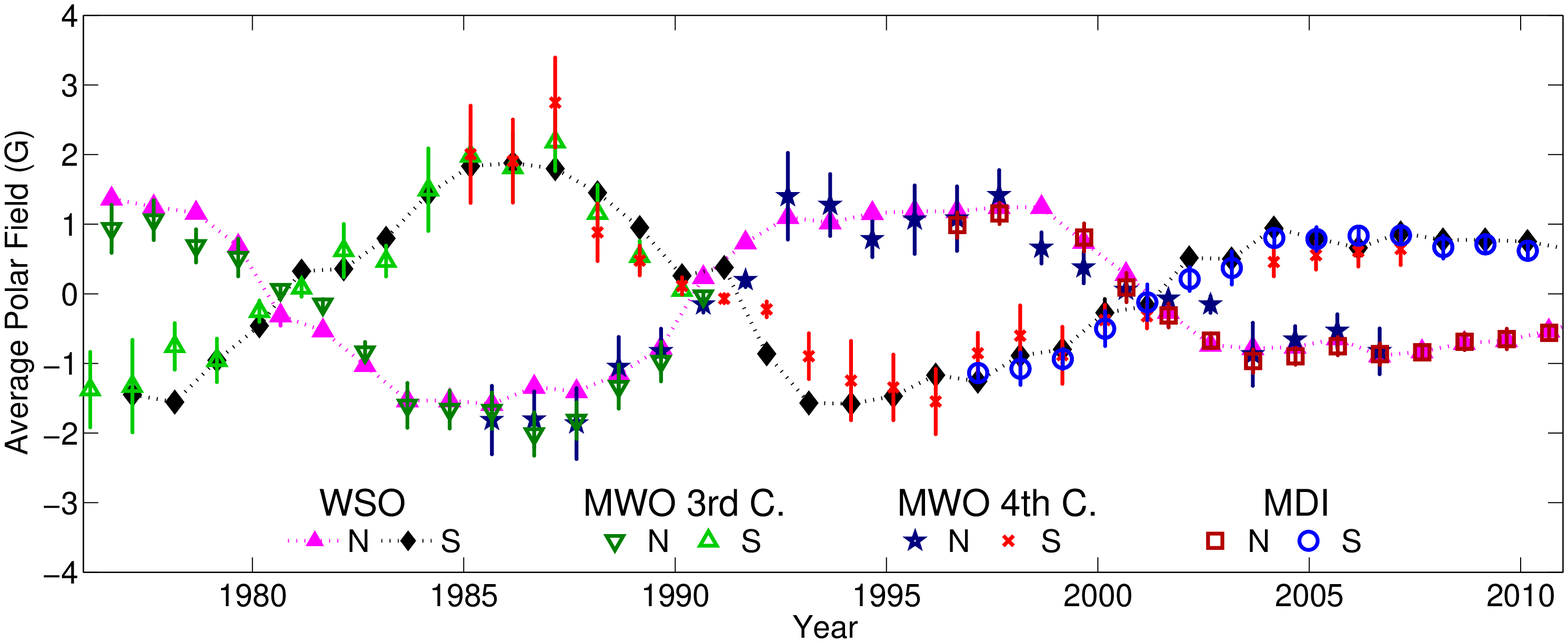}\\
                        (c)
\end{tabular}
\caption{(a) Scatter plot comparing MDI average polar field with WSO measurements of the polar field.  (b) Scatter plot comparing MWO facular measurements with WSO measurements of the polar field.  Each scatter plot is fitted with a line passing through the origin (dot-dashed red) and the 95\% confidence interval is bounded by dotted green lines.  (c) Overplot of the WSO polar field measurements (dotted lines plus solid triangles-NP, diamonds-SP), the MWO calibrated facular measurements of the 3rd campaign (open downward-NP, upward-SP triangles), the 4th campaign (stars-NP, crosses-SP), and the corrected MDI average polar field (open squares-NP, circles-SP).  MWO facular measurements are converted to polar field strength using a multiplication factor of $0.08\pm0.01$ G and the MDI average polar field is multiplied by a factor of $0.53\pm0.03$.  Both correspond to the slopes and the 95\% confidence interval of their respective linear fits (panels a and b).}\label{Fig_MWO_WSO_MDI}
\end{figure*}

As shown in Figure \ref{Fig_MWO_WSO_MDI}-a, polar field measurements taken by the WSO agree very well with the average polar field calculated using MDI.  Though values are not identical (WSO measurements are roughly half MDI averages), the difference can be easily corrected through a multiplicative factor and the relationship between them is fit very well by a line that passes through the origin (with a goodness of fit equal to 0.97; see Figure \ref{Fig_MWO_WSO_MDI}-a).  A similar result is obtained when comparing WSO magnetic field measurements and MWO facular count, whose relationship is also fit well by a line that passes through the origin (with a goodness of fit equal to 0.90; see Figure \ref{Fig_MWO_WSO_MDI}-b).  The general agreement between these quantities is evident in Figure \ref{Fig_MWO_WSO_MDI}-c, which shows their overlap after being corrected by the factors obtained by the linear fits (Figures \ref{Fig_MWO_WSO_MDI}-a \& b). In Figure \ref{Fig_MWO_WSO_MDI}-c WSO measurements are joined with a dotted line indicating that they are the reference dataset.  The very good agreement between the WSO magnetic field measurements and the facular data belonging to two different MWO data reduction campaigns (after calibration) shows the direct relationship between facular and magnetic data, and helps validate the calibration process used to standardize the different MWO data reduction campaigns.

\subsection{Polar Faculae as a Proxy for Polar Signed Magnetic Flux}\label{Sec_PFlx}

Although there is no doubt about the direct relationship between facular measurements and polar field strength, it is useful to study the association with magnetic flux since it is a more relevant quantity for a wide range of studies such as the prediction of the solar cycle amplitude (Choudhuri, Chatterjee \& Jiang 2007\nocite{choudhuri-chatterjee-jiang07}), modulation of cosmic ray intensity at Earth (Usoskin et al.\ 2002\nocite{usoskin-etal02}; Solanki, Sch\"ussler \& Fligge 2000\nocite{solanki-schussler-fligge00}; Cliver, Richardson \& Ling 2011\nocite{cliver-richardson-ling11}), and the understanding of solar wind properties (Luhmann et al.\ 2009\nocite{luhmann-etal09}; Wang, Robbrecht \& Sheeley 2009\nocite{wang-robbrecht-sheeley09}).  In this section we use MDI LOS magnetograms as the reference dataset because their spatial resolution allows the calculation of signed polar flux.  This means that in terms of signed polar flux, both the MWO polar facular measurements and the WSO LOS magnetic field measurements are calibrated to their MDI counterparts.  We calculate the total signed polar flux as follows:
\begin{enumerate}
  \item Convert the MDI LOS magnetic field into a radial field assuming that this is the only component of the magnetic field at the surface.  To do this we divide the LOS magnetic field by the cosine of the angle between the vector normal to the surface and the vector pointing towards the observer.
  \item Calculate the area of each pixel taking account of the projection effect due to its relative position on the Sun with respect to the observer.
  \item Integrate the flux associated with each observed pixel ($B_r A_p$) for all latitudes poleward of  $70^o$.  The outer three radial pixels on the solar disk (corresponding to an angular size of 6") are ignored in order to reduce the effect of noise (enhanced by the LOS correction) on the polar flux calculation.
\end{enumerate}

\begin{figure*}
\centering
\begin{tabular}{c}
  \includegraphics[scale=0.43]{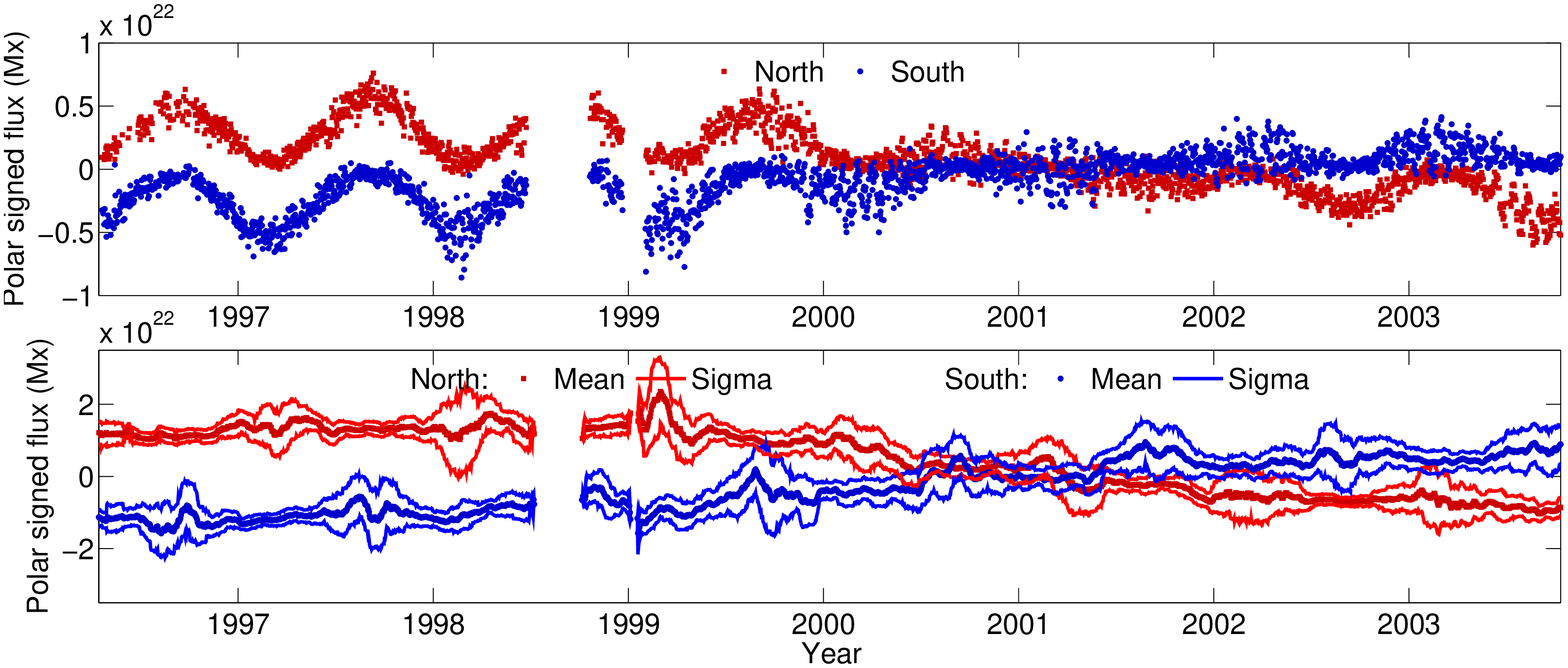}\\
  (a)\\
  \includegraphics[scale=0.43]{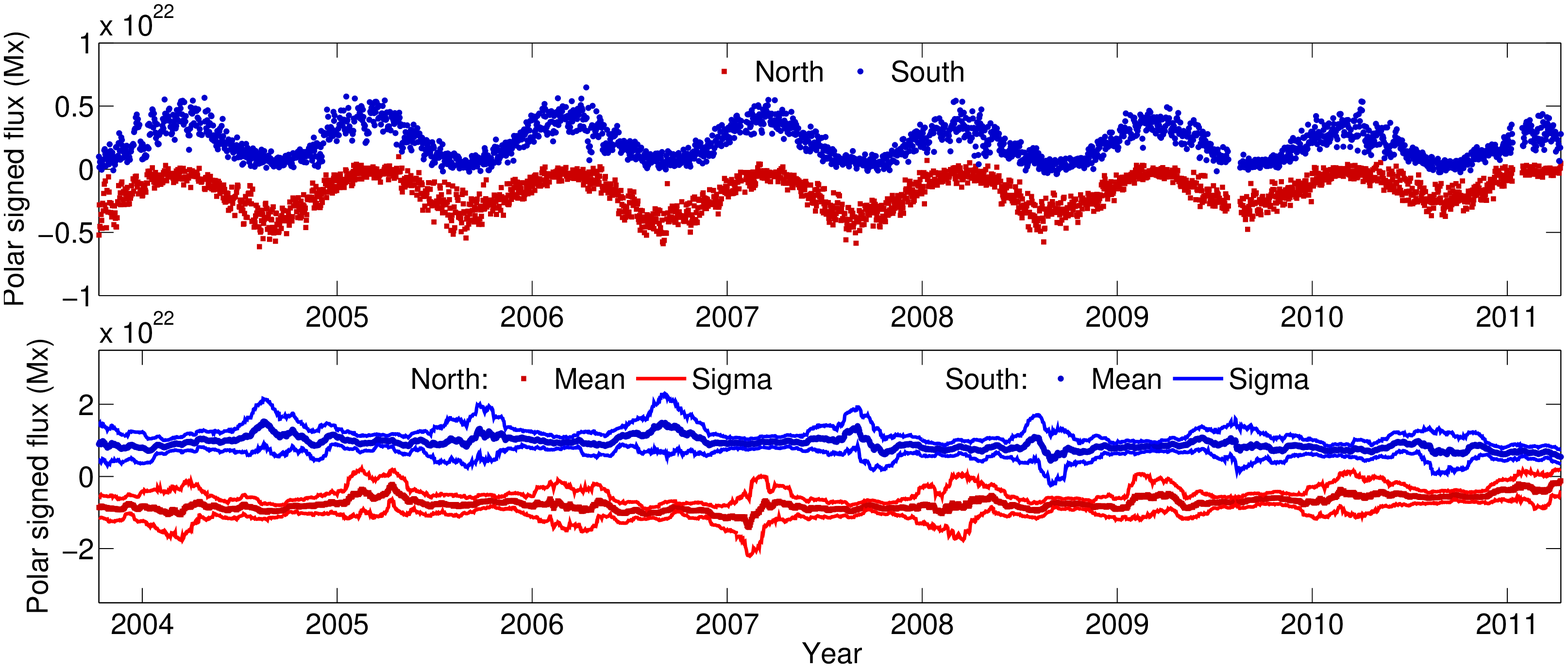}\\
  (b)\\
  \includegraphics[scale=0.43]{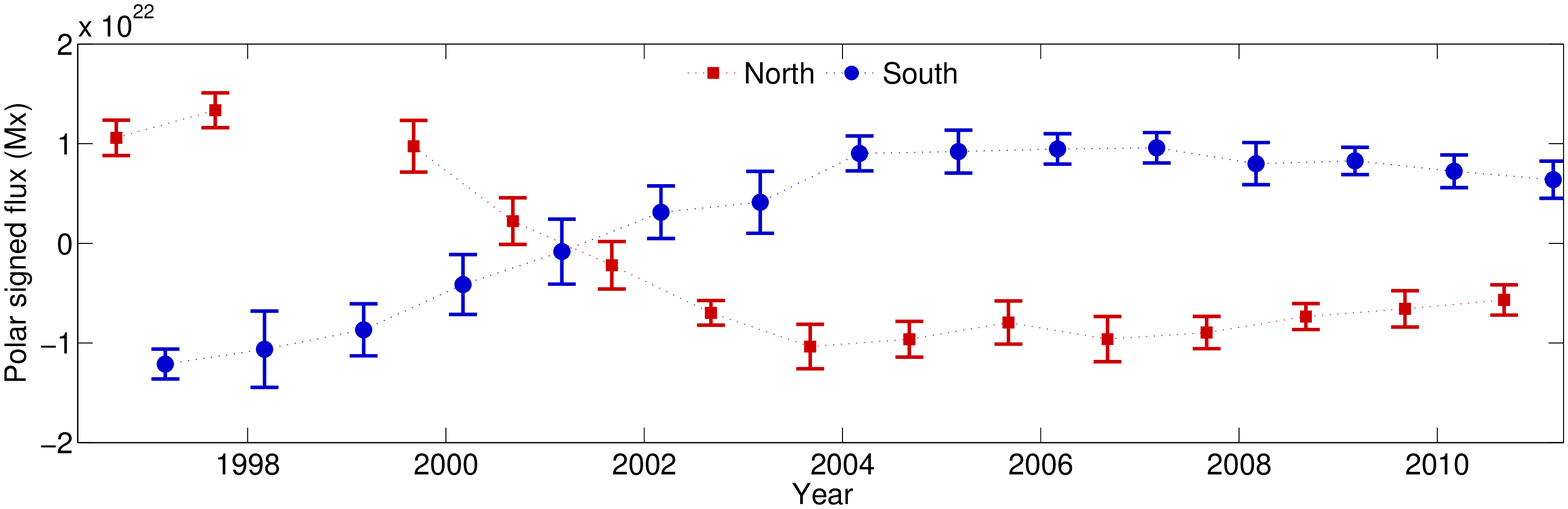}\\
  (c)
\end{tabular}
\caption{MDI total signed polar flux above (below) $70^o$ ($-70^o$) for the 1996-2004 (a) and 2004-2011 (b) intervals.  The top panel of each figure shows the observed daily polar signed flux; the bottom panel shows polar flux after it has been corrected for the unseen backside area and then smoothed using a month-long running average.  Red (blue) corresponds to the north (south) pole and the thin lines correspond to one standard deviation.  (c) Selected values of the average signed polar flux for the 1996-2011 interval.  Points corresponding to the 4-Mar (4-Sep) of each year are used for the south (north) pole.  These dates correspond to the point in Earth's orbit with the best observations of each particular pole.}\label{Fig_MDI_Flux_Pros}
\end{figure*}

The top panels of Figures \ref{Fig_MDI_Flux_Pros}-a \& b show the daily values of calculated polar flux.  Like the facular count (Figs.~\ref{Fig_Fac_D}-a \& n) and average polar field (Figs.~\ref{Fig_MDI_LOS_Pros}-a \& b), there is a clear yearly modulation of polar flux due to the change in observable area poleward of $70^o$. The modulation is significantly larger in the case of signed polar flux.  In order to remove this annual variation, we correct each daily polar flux measurement using a factor equal to the total area of the polar cap divided by the visible area.  As noted above the polar cap is defined as the area poleward of $70^o$.  Finally, we perform a month-long running average in order to remain consistent with the time intervals used for other quantities.  The result (shown in the bottom panels of Figs.~\ref{Fig_MDI_Flux_Pros}-a \& b), has a much weaker yearly modulation though it shows clear ``pulses" in the standard deviation corresponding to the periods in which the pole in question is barely visible from Earth.

\begin{figure*}
\centering
\begin{tabular}{c}
\begin{tabular}{cc}
  \includegraphics[scale=0.4]{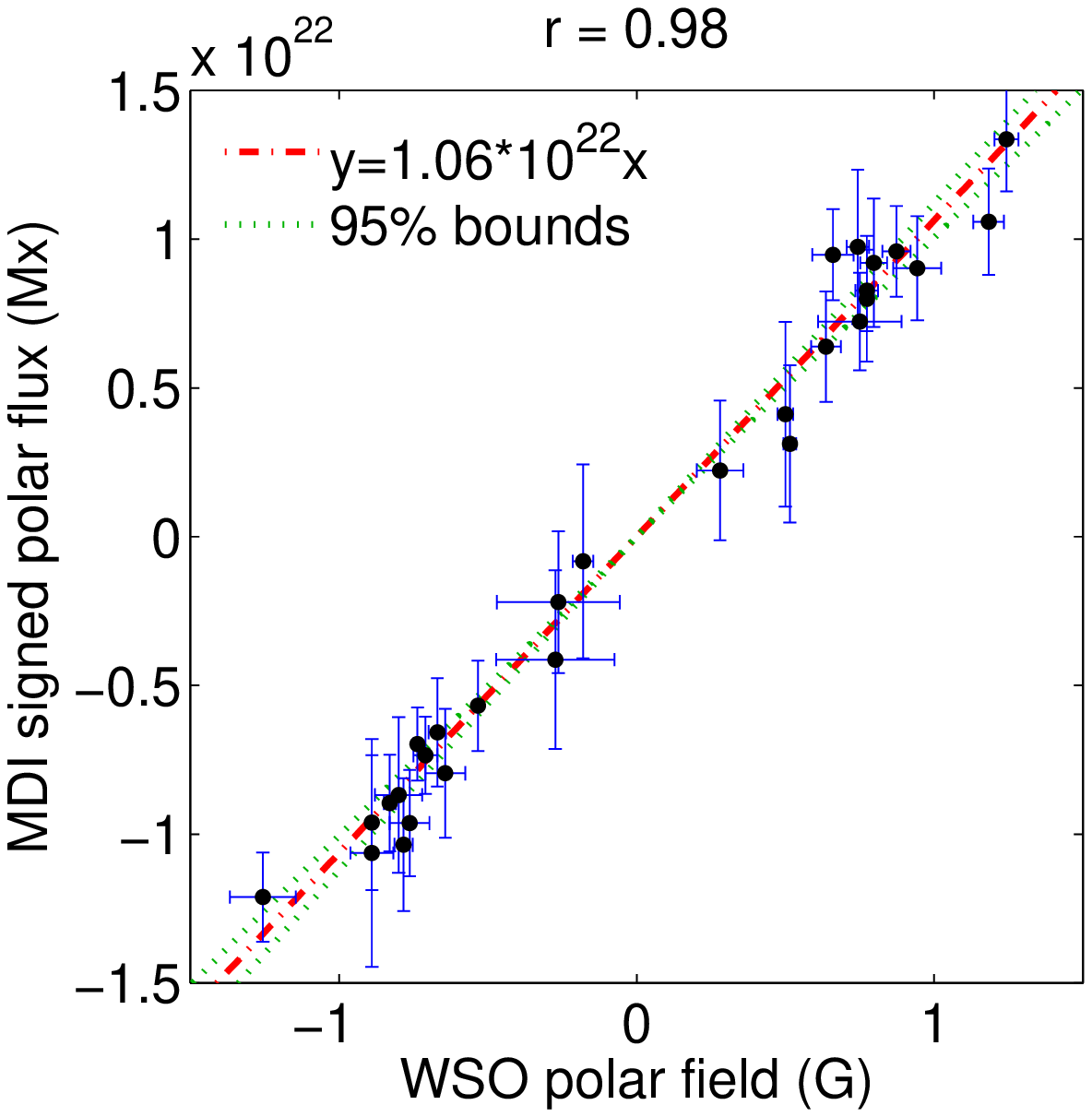} & \includegraphics[scale=0.4]{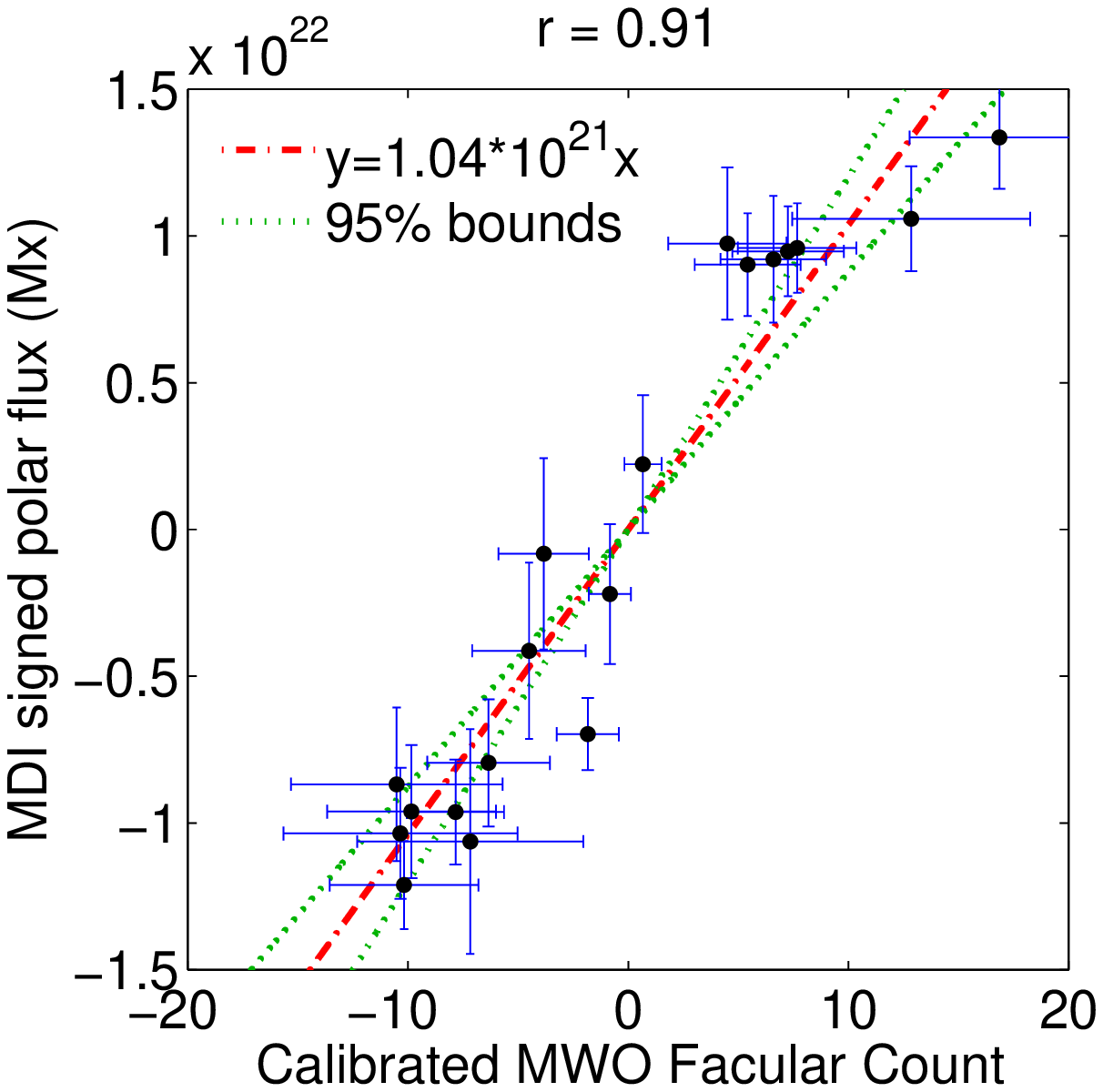} \\
                    (a)                              &                   (b)
\end{tabular}\\
  \includegraphics[scale=0.35]{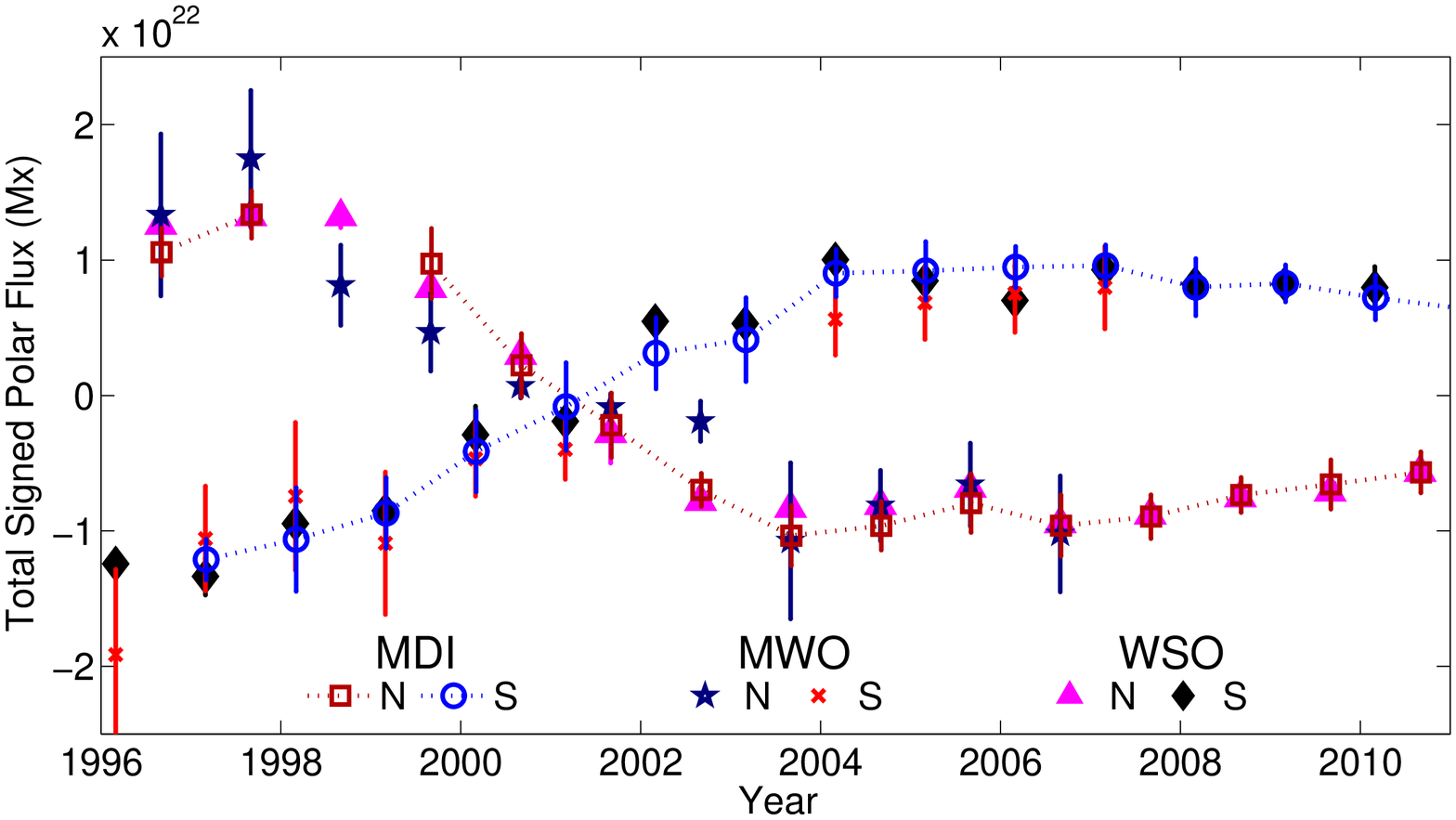}\\
                        (c)
\end{tabular}
\caption{(a) Scatter plot comparing WSO measurements of the polar field with MDI signed polar flux.  (b) Scatter plot comparing MWO facular measurements with MDI signed polar flux.  Each scatter plot is fitted with a line passing through the origin (dot-dashed red) and the 95\% confidence interval is bounded by dotted green lines.  (c) Overplot of the MDI signed polar flux (dotted lines plus open squares-NP, circles-SP), MWO calibrated facular measurements (stars-NP, crosses-SP), and WSO polar field measurements (solid triangles-NP, diamonds-SP).  MWO facular measurements are converted to polar field strength using a multiplication factor of $(1.04\pm0.16)\times10^{21}$Mx and the WSO polar field measurements are multiplied by a factor of $(1.06\pm0.06)\times10^{22}$Mx/G.  Both correspond to the slopes and the 95\% confidence interval of their respective linear fits (panels a and b).}\label{Fig_MWO_WSO_MDI2}
\end{figure*}

In the usual way, we select values of polar magnetic flux corresponding to 4-Mar (4-Sep) of each year for the south (north) pole (see Figure \ref{Fig_MDI_Flux_Pros}-c) which we can use to compare to MWO and WSO measurements.  Once again we find good agreement between MDI signed polar flux, WSO magnetic field measurements (goodness of fit 0.98; Figure \ref{Fig_MWO_WSO_MDI2}-a), and MWO facular count (goodness of fit 0.91; Figure \ref{Fig_MWO_WSO_MDI2}-b).  This agreement is also evident in an overplot of the three quantities, shown in Figure \ref{Fig_MWO_WSO_MDI2}-c, where MDI measurements are joined with a dotted line indicating the fact that they are the reference dataset.

\begin{figure*}
\centering
  \includegraphics[scale=0.525]{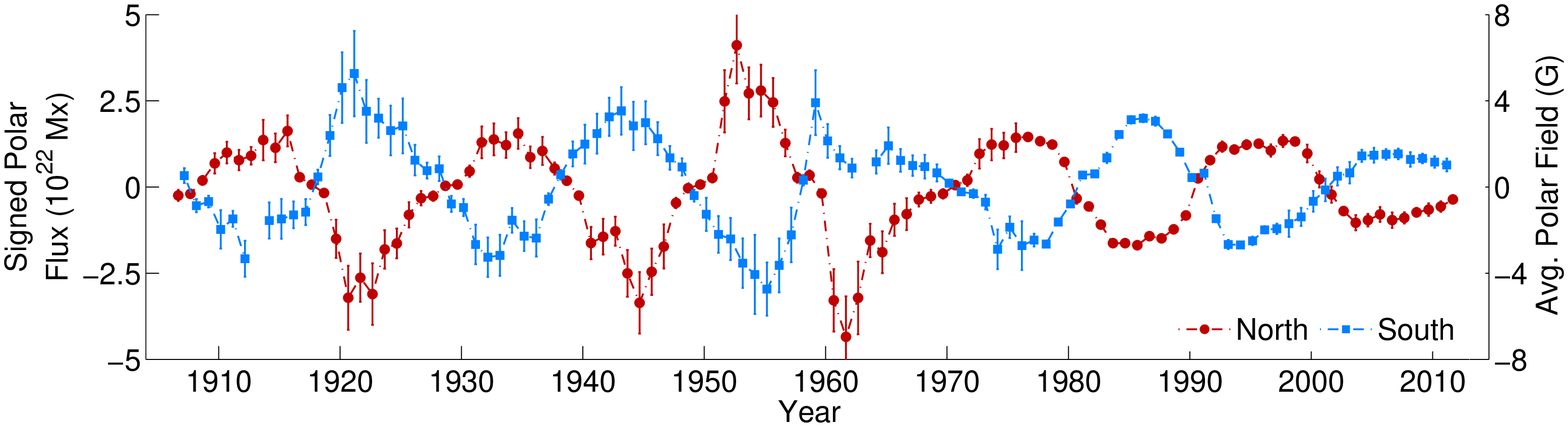}
  \caption{Consolidated signed polar flux database referenced to MDI measurements.  Data-points were taken from MDI (1996-2010), WSO (1975-1996), and MWO (1906-1975).  The second axis shows the equivalent values of average polar field strength referenced to WSO measurements.}\label{Fig_Cons}
\end{figure*}

After the calibration and validation of the polar faculae database (using data from MDI, WSO and MWO), we are ready to produce a consolidated database for the evolution of the polar magnetic flux since 1906.  We use the different calibration factors mentioned in previous sections (referenced to MDI) and select data from MDI (1996-2010), WSO (1975-1996), and MWO (1906-1975); the resultant data series is shown in Figure \ref{Fig_Cons}.  It is important to mention that due to the necessary propagation of errors incurred in cross-calibrating the MWO data, care must be taken when comparing polar flux values before and after 1975 (as can be observed in Figure \ref{Fig_Cons} from the relative size of the errorbars).  However, such a comparison can be done if one allows for a 7\% error in this section of the database.

\section{The Role of Polar Magnetic Flux in Determining the Properties of the Heliospheric Magnetic Field}


It has been found that one of the most important quantities when determining the evolution of geomagnetic activity indicators (which have been measured for more than a century) is the Heliospheric Magnetic Field (HMF; Stamper et al.\ 1999\nocite{stamper_etal99}).  Direct HMF observations exist since the beginning of 1965 and these observations can be extended all the way to the mid 1800's using the \emph{aa} geomagnetic index (Lockwood, Rouillard \& Finch 2009\nocite{lockwood_rouillard_finch09}) and the InterDiurnal Variability geomagnetic index (Svalgaard \& Cliver 2010\nocite{svalgaard_cliver10}).  Following a decade of vigourous debate, different reconstructions of HMF based on geomagnetic data have gradually reached consensus (see Lockwood \& Owens 2011\nocite{lockwood-owens11}, Svalgaard \& Cliver 2010\nocite{svalgaard_cliver10}, and references therein), providing us with the opportunity of studying the combined role of active regions (ARs) and polar flux in determining the characteristics of the heliospheric environment (as well as acting as a consistency check for our polar flux database).

Due to the rapid decay of high order multipoles with distance, the HMF strongly determined by the axial and equatorial dipolar components of the photospheric field.  Surface flux-transport simulations (Wang, Lean \& Sheeley 2005\nocite{wang-lean-sheeley05}) have shown that the axial dipole is strongly correlated to the polar flux and the equatorial dipole is strongly correlated to the square root of the sunspot number.   However, the lack a of sufficiently long proxy for the evolution of the polar magnetic field has made it difficult to study this relationship in detail.  In this section we will address this issue using HMF measurements taken from the OMNI dataset (\href{http://omniweb.gsfc.nasa.gov/}{http://omniweb.gsfc.nasa.gov/}), the HMF reconstructions of Lockwood, Rouillard \& Finch (2009\nocite{lockwood_rouillard_finch09}), and Svalgaard \& Cliver (2010\nocite{svalgaard_cliver10}), updated sunspot area data of Balmaceda et al.\ (2009\nocite{balmaceda-etal09}; sunspot area acts as a better proxy for AR flux than sunspot number does, see Dikpati, de Toma \& Gilman 2006\nocite{dikpati-detoma-gilman06}; Solanki \& Schmidt 1993\nocite{solanki_schmidt93}), and the average of the amplitude of the polar magnetic flux in both hemispheres.


\begin{figure*}
\centering
  \includegraphics[width=0.76\textwidth]{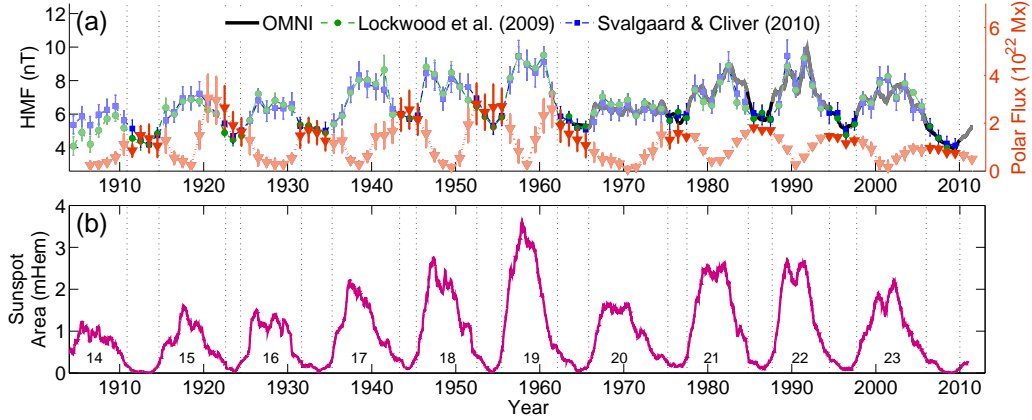}
  \caption{(a) Overplot of the HMF taken from the OMNI database (black solid line), the reconstructed HMF of Lockwood, Rouillard \& Finch (2009; green dashed line with circles), Svalgaard \& Cliver (2010; blue dot-dashed line with squares) and the average of the polar flux amplitude in both hemispheres (orange dotted line with triangles).  (b) Yearly smoothed total sunspot area of Balmaceda et al. (2009). Vertical dashed lines bracket intervals corresponding to sunspot minima.  In the top panel colors are also accentuated during minima.}\label{Fig_HMF_PF_SA}
\end{figure*}

The correlation of both sunspot area and polar flux with HMF is evident in Figure \ref{Fig_HMF_PF_SA} where we show an overplot of HMF measurements and reconstructions with polar magnetic flux (Fig.~\ref{Fig_HMF_PF_SA}-a) along with a yearly smoothed total sunspot area (Fig.~\ref{Fig_HMF_PF_SA}-b).  The figure shows clearly how HMF amplitude rises and falls with the sunspot cycle, falling to values dictated by the average polar flux during solar minima (denoted by bracketing black dotted lines).

\begin{figure*}
\centering
\begin{tabular}{cc}
  \includegraphics[width=0.3\textwidth]{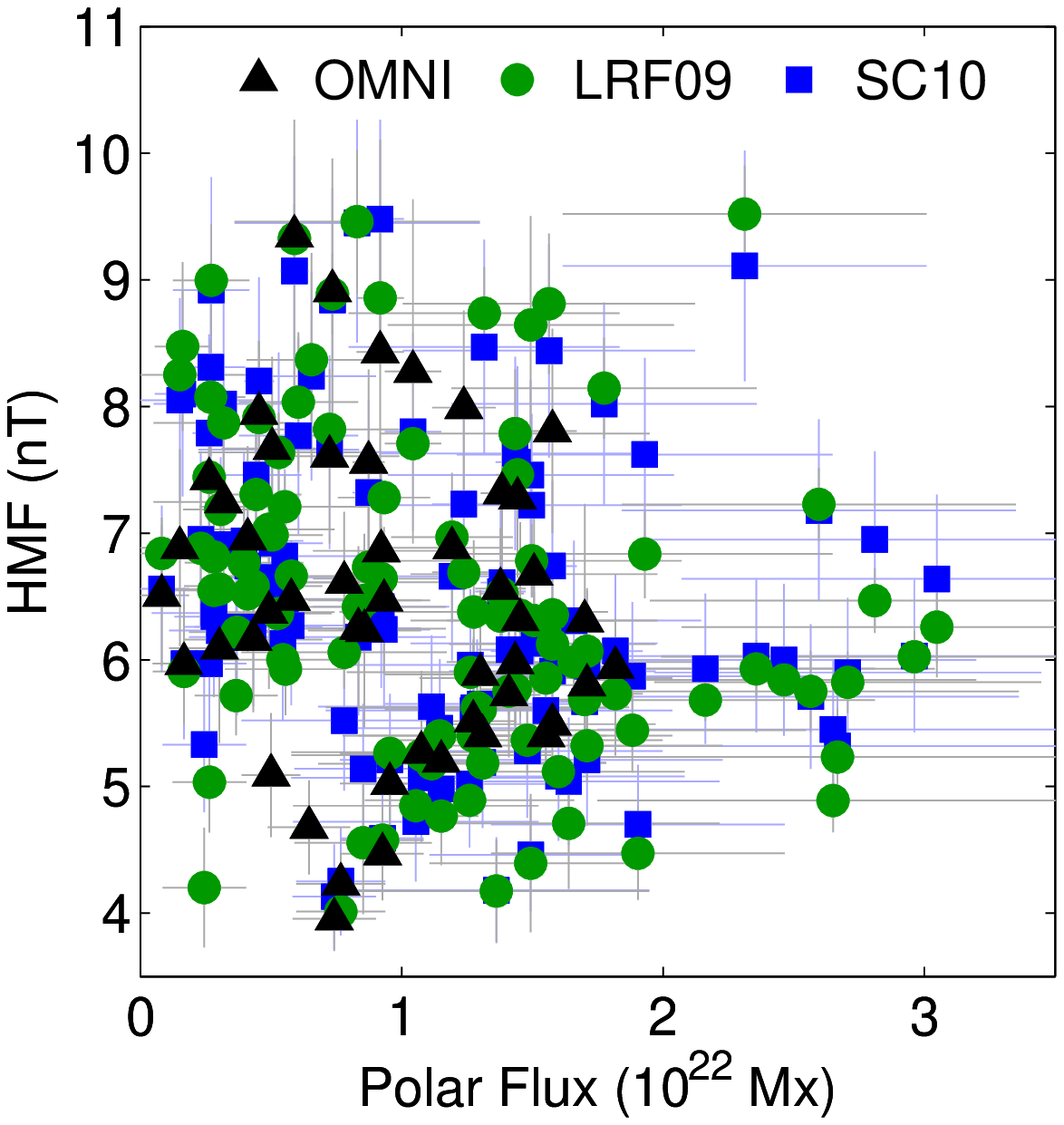} & \includegraphics[width=0.3\textwidth]{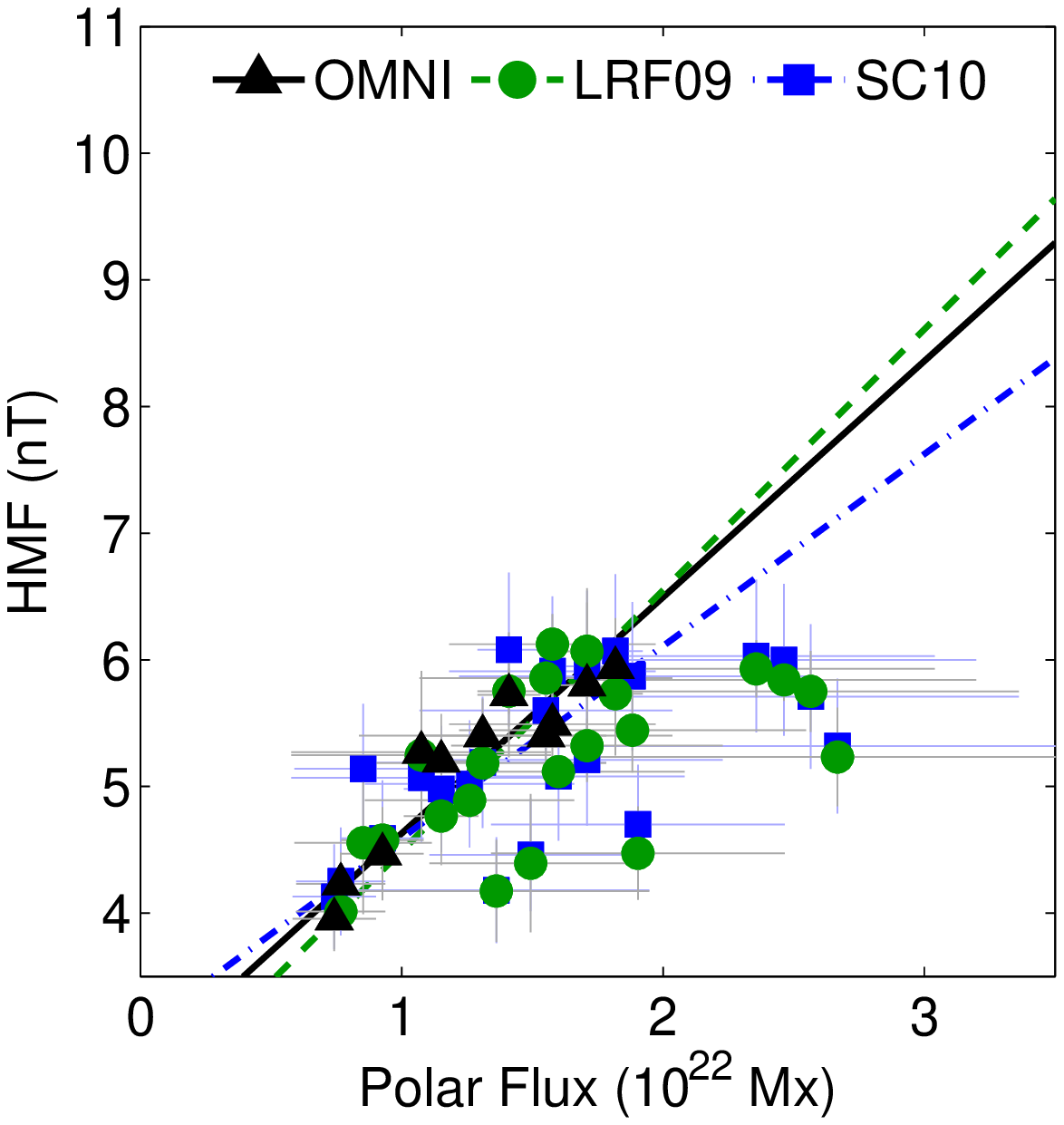}\\
  (a) & (b)
\end{tabular}
\caption{Scatter-plot of HMF vs.\ polar flux.  Using all measurements shows no clear correlation (a).  The correlation arises by limiting the plot to points at solar minimum (b).   Weighted total least squares fits for all datasets yields similar relationships -- mainly determined by data-points taken after 1975 (the points with the smallest errors), but consistent with the errorbars of all measurements.  The Spearman's rank correlation coefficient or the reconstructed HMF of Lockwood, Rouillard \& Finch ( Svalgaard \& Cliver) is $r=0.54$ ($r=0.63$) with $P=99\%$ ($P=99\%$) confidence level.  The Spearman's rank correlation coefficient for OMNI HMF measurements at minimum is $r=0.96$ with $P=99\%$ confidence level.  }\label{Fig_CorrP}
\end{figure*}

The relationship between the polar flux at solar minimum and HMF is illustrated in Figure \ref{Fig_CorrP} where we compare a scatter-plot obtained using all yearly measurements (Fig.~\ref{Fig_CorrP}-a) with one restricted to measurements taken during solar minimum (Fig.~\ref{Fig_CorrP}-b).  We perform a linear fit of the form:
\begin{equation}\label{Eq_PFM}
    HMF_{min} = a_0 + a_1 PF,
\end{equation}
using a Weighted Total Least Squares (WTLS) algorithm  in order to incorporate errorbar information into the fit (Krystek \& Anton 2007\nocite{krystek-anton07}), and obtain comparable relationships (plotted as lines in Fig.~\ref{Fig_CorrP}-b) for OMNI HMF data and both HMF reconstructions.  We find a good agreement between the fits and the data (see Table \ref{Tab_FitsPFM}). The fit is particularly good for the OMNI data, whose time-span of direct HMF measurements coincides with direct measurements of polar field taken by the WSO.
Further evidence supporting this claim can be obtained from calculating the Spearman's rank correlation coefficient (Spearman 1904\nocite{spearman1904}; see Table \ref{Tab_FitsPFM}), which in the case of the OMNI data is $r=0.96$ with $P=98\%$ confidence level.

\begin{table*}
\begin{center}
\footnotesize
\begin{tabular*}{\textwidth}{@{\extracolsep{\fill}}  l c c c c c c}
\multicolumn{7}{c}{\normalsize \textbf{Fits of Polar Flux During Minimum to HMF}}\\
\toprule
                           & \multicolumn{4}{c}{\textbf{Weighted Total Least-Squares Fit}} & \multicolumn{2}{c}{\textbf{Spearman's Correlation}}\\
                           \cmidrule{2-5}                          \cmidrule{6-7}
\textbf{HMF dataset}       & $a_0$ (nT) & $a_1$ ($10^{-22}$nT/Mx) & $\sqrt{\chi_\nu^2}$ & Significance & $r$-Coefficient & Significance\\
\midrule
OMNI dataset               & 2.77 & 1.86 & 0.46 & $99\%$ & 0.96 & $99\%$ \\
Lockwood et al.\ (2009)    & 2.43 & 2.06 & 0.88 & $73\%$ & 0.54 & $99\%$\\
Svalgaard \& Cliver (2010) & 3.08 & 1.52 & 0.77 & $92\%$ & 0.63 & $99\%$\\
\bottomrule
\end{tabular*}
\end{center}
\hspace{1em}
  \caption{Fit parameters of polar flux during minimum (Eq.\ \ref{Eq_PFM}) to the different HMF databases used in this work.  The fit is done using a Weighted Total Least-Squares (WTLS) algorithm which minimizes $\chi^2$ error.  The reduced $\chi^2$ error ($\sqrt{\chi_\nu^2}$) and its statistical significance are used to estimate goodness-of-fit (a value less than or equal to 1 indicates a good fit).  Additionally, we calculate the Spearman's rank correlation coefficient and find a good correlation (values close to 1 indicate a very strong correlation), specially for measurements taken after 1965 (OMNI data).}\label{Tab_FitsPFM}
\end{table*}

To evaluate the relationship between HMF and sunspot area (as a proxy for AR flux) we make a least-squares fit of sunspot area, and the square root of sunspot area to both the reconstructed and measured HMF databases:
\begin{equation}\label{Eq_SAM}
    HMF = a_0 + a_1 SA,
\end{equation}
\begin{equation}\label{Eq_SAM2}
    HMF = a_0 + a_1 \sqrt{SA},
\end{equation}
where $SA$ corresponds to sunspot area normalized by its maximum value.  We evaluate the performance of these fits by comparing their Root Mean Square Error (RMSE), which is a measure of the average error between a least-squares fit and the fitted data (the smaller the better).   As can be observed in Table \ref{Tab_FitsSA}, for all HMF data sources, the square root of sunspot area yields a tighter fit than the linear sunspot area (i.e.  Eq.~\ref{Eq_SAM2} performs better).  This is due to the fact that ARs erupt at random longitudes, causing the equatorial phases of the equatorial dipole moments to be random as well (Wang \& Sheeley 2003\nocite{wang-sheeley03}).

\begin{figure*}
\centering
  \includegraphics[width=0.76\textwidth]{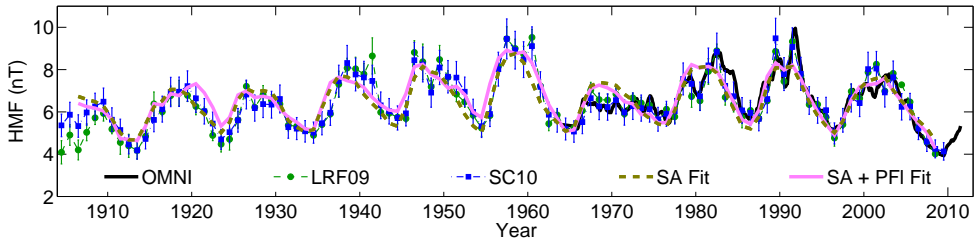}
  \caption{Overplot of the HMF taken from the OMNI database (black solid line), the reconstructed HMF of Lockwood, Rouillard \& Finch (2009; dashed line with circles), Svalgaard \& Cliver (2010; dot-dashed line with squares), the least-squares fit to the square root of sunspot area (thick dashed line) and the least-squares fit to a linear combination of the square root of sunspot area and the average of the polar flux amplitude in both hemispheres (thick solid magenta line).}\label{Fig_Fits}
\end{figure*}

Finally, we combine both sunspot area and polar flux (as a proxy for the axial dipolar moment of the Sun's magnetic field) in a combined fit:
\begin{equation}\label{Eq_SAPFM}
    HMF = a_0 + a_1 \sqrt{SA} + a_2 PF,
\end{equation}
where PF corresponds to the average of the amplitude of the polar magnetic flux in both hemispheres normalized by its maximum value.  It can be observed in in Table \ref{Tab_Fits} that, in every case, the introduction of polar flux into the fit further reduces RMSE and the fit more closely tracks the observed data, specially evident during the extended minimum of cycle 23; see Figure \ref{Fig_Fits}.   However, to ensure that polar and AR flux (characterized by sunspot area) have more explanatory power together (than sunspot area alone),  we perform an F-test using the reduced and extended models (Eqs.~\ref{Eq_SAM2} vs.~\ref{Eq_SAPFM}).   This way we make sure the improvement in the fit is not simply caused by the increase in fitting parameters.  Our null hypothesis is that $a_2 = 0$, or in other words, that the extended model does not provide a significantly better fit than the reduced model.  Our test statistic is:
\begin{equation}\label{Eq_Ftest}
    F = \frac{(RSS_r - RSS_e)/(n_e - n_r)}{RSS_e/(N-n_e)},
\end{equation}
where $RSS_r$ and $RSS_e$ are the residual sum of squares ($RSS = \sum_i^n (y_i-f(x_i))^2$) associated with the reduced ($r$) and extended ($e$) models, $n_e$ and $n_r$ are the amount of fitting parameters in each model, and $N$ is the total amount of fitting points.  This test statistic quantifies the relationship between the relative decrease in RSS and the relative increase in degrees of freedom caused by the introduction of additional fitting parameters.  If the null hypothesis is correct, the test statistic has an F distribution with $(n_e - n_r)$ and $(N-n_e)$ degrees of freedom (for more details about the F-test please refer to Snedecor \& Cochran 1989\nocite{snedecor_cochran89}).

Our system is described by the following numbers: for our reduced and extended models $n_r = 2$ and $n_e = 3$; for the HMF reconstructions $N = 103$ and for OMNI data $N = 48$.   This means that the null hypothesis can be rejected with a statistical significance of $99\%$ if F is greater or equal than $6.89$ ($7.23$) in the case of the HMF reconstructions (OMNI data).  Considering that these values are below the F test statistic calculated for the different fits shown in Table \ref{Tab_Fits} (the lowest of which is $14.43$), allows us to reject the null hypothesis and show that polar flux and sunspot area (as a proxy for AR flux) have more explanatory power together in terms of the time evolution of HMF than AR flux alone does.   It is important to note that the performed fits and statistical tests have the added benefit of acting as a sanity check for our calibration across different datasets (MDI, WSO, and MWO) and different data reduction campaigns.  However, more important is the verification of the theoretical relationship between one of the quantities most relevant at determining the conditions of the near-Earth environment (HMF) and those directly connected with the long-term evolution of the solar magnetic field.  The further refining of this relationship, along with the improvement our models for the evolution of the solar magnetic field, can only improve our understanding the Sun-Earth connection and enhance our space-weather (and space-climate) forecasting capabilities.

\begin{table}
\begin{center}
\footnotesize
\begin{tabular*}{0.5\textwidth}{@{\extracolsep{\fill}}  l c c c}
\multicolumn{4}{c}{\normalsize \textbf{Fits of $SA$ and $\sqrt{SA}$ to HMF}}\\
\toprule
\textbf{HMF dataset}       & $a_0$ & $a_1$ & RMSE (nT)\\
\midrule
{\normalsize\textbf{Fits of $SA$ to HMF}}\\
OMNI dataset               & 5.32  & 2.93  & 0.70\\
Lockwood et al.\ (2009)    & 5.14  & 4.39  & 0.73\\
Svalgaard \& Cliver (2010) & 5.22  & 4.23  & 0.63\\
\midrule
{\normalsize\textbf{Fits of $\sqrt{SA}$ to HMF}}\\
OMNI dataset               & 4.43  & 3.60  & 0.65\\
Lockwood et al.\ (2009)    & 4.12  & 4.69  & 0.70\\
Svalgaard \& Cliver (2010) & 4.24  & 4.51  & 0.61\\
\bottomrule
\end{tabular*}
\end{center}
\hspace{1em}
  \caption{Fit parameters of sunspot area (Eqs.~\ref{Eq_SAM} \& \ref{Eq_SAM2}) to the different HMF databases used in this work.  The top part of the table corresponds to fits of sunspot area (Eq.~\ref{Eq_SAM}), the bottom part corresponds to fits of the square roots of sunspot area (Eq.~\ref{Eq_SAM2}).   RMSE corresponds to the Root Mean Square Error (also referred to as Root Mean Square Deviation), which is a measure of the average error between a least-squares fit and the fitted data (the smaller this quantity is, the better the fit is).}\label{Tab_FitsSA}
\end{table}

\begin{table*}
\begin{center}
\footnotesize
\begin{tabular*}{\textwidth}{@{\extracolsep{\fill}}  l c c c c c c c c}
\multicolumn{9}{c}{\normalsize \textbf{Fits of $\sqrt{SA}$ and $PF$ to HMF}}\\
\toprule
                           & \multicolumn{3}{c}{\textbf{Sunspot Area Only}} & \multicolumn{4}{c}{\textbf{Sunspot Area + Polar Flux}} & \textbf{F-Test} \\
\textbf{HMF dataset}       & $a_0$ & $a_1$ & RMSE (nT)            & $a_0$ & $a_1$ & $a_2$ & RMSE (nT)               & \\
\midrule
OMNI dataset               & 4.43  & 3.60  & 0.65                 & 2.74  & 4.70  & 2.03  & 0.49                    & 33.91\\
Lockwood et al.\ (2009)    & 4.12  & 4.69  & 0.70                 & 3.33  & 5.34  & 1.22  & 0.66                    & 14.43\\
Svalgaard \& Cliver (2010) & 4.24  & 4.51  & 0.61                 & 3.40  & 5.20  & 1.31  & 0.55                    & 24.30\\
\bottomrule
\end{tabular*}
\end{center}
\hspace{1em}
  \caption{Fit parameters of the square root of sunspot area (left group, Eqs.~\ref{Eq_SAM}) and the quare root of sunspot area with polar flux (middle group, Eqs.~\ref{Eq_SAPFM}) to the different HMF databases used in this work.  RMSE corresponds to the Root Mean Square Error (also referred to as Root Mean Square Deviation), which is a measure of the average error between a least-squares fit and the fitted data (the smaller this quantity is, the better the fit is).  The rightmost column shows the values of the F-test used to evaluate whether polar flux and sunspot area have more explanatory power than sunspot area alone; any value of F above $6.89$ ($7.23$) in the case of the HMF reconstructions (OMNI data) permits us to reject the null hypothesis (with a statistical significance of 99\%) that the improvement of the larger model (fitting of both the square root of sunspot area and polar flux) is only caused by the increase in fitting parameters.  Note that the null hypothesis can be rejected in every case.}\label{Tab_Fits}
\end{table*}

\section{Concluding Remarks}

The focus of this work has been the standardization, validation and calibration of a long-term facular dataset.  For the first part we take advantage of the overlapping intervals across the four different MWO data reduction campaigns and show how, in spite of some underestimation in facular count from one campaign to the next, the discrepancies can be removed by the application of a multiplicative factor (leaving us with a self-consistent database spanning more than a hundred years of observations).

Our second goal was to validate the facular database using two different approaches:  the first one is to compare the MWO dataset with an MDI facular count obtained through an automatic detection algorithm; the excellent agreement between MDI and WSO facular count testifies to the validity of the methodology used to count MWO faculae (independently of the fact that it is based on eye estimates).   We also compare the MWO standardized facular count belonging to different data reduction campaigns with the LOS magnetic field measurements taken by the WSO.  The correspondence between them argues a strong case in favor the adequacy of multiplicative factors as means of standardizing the entire MWO facular dataset.

Given that our ultimate goal is to use MWO facular count as a magnetic proxy, we combine it with MDI total signed polar flux in order to obtain a calibration factor of $(1.04 \pm 0.16)\times10^{21}$Mx/facula for the standardized facular dataset.   It is important to note that this calibration factor does not mean that each polar faculae contains $10\times10^{21}$Mx, but rather that if one wants to convert the standardized polar facular count into MDI total signed polar flux this is the factor one should use.  This calibration factor can in turn be compared with flux measurements of low-latitude faculae as they transit the solar disk (Sheeley 1964, 1966).   However, it must be referenced first to the facular values of the first MWO data reduction campaign (since the standardized facular dataset is referenced to the third MWO data reduction campaign and the MWO estimate was performed during the first data reduction campaign; Sheeley 1964, 1966).   When converted to facular values of the first MWO campaign we obtain a factor of $(0.66 \pm 0.20)\times10^{21}$Mx/facula, which matches well the MWO factor of $(0.34 \pm 0.14)\times10^{21}$Mx/facula (Sheeley 1966\nocite{sheeley66}).  The reason for the slight overestimation resides in the fact that the MWO estimate is limited exclusively to facular regions whereas we use the entire polar signed flux in our estimate.  However, this agreement is highly encouraging because it essentially means that most of the signed flux present in the poles is indeed associated with polar faculae and thus facular count can be used as an excellent proxy of what the poles were doing during more than a hundred years.

Our next step is to combine our polar flux database with the sunspot area database of Balmaceda et al.\ (2009\nocite{balmaceda-etal09}) in order to study the role of the polar flux in the evolution of the heliospheric magnetic field.  To do so we perform different fits to the HMF taken from OMNI data, as well as the HMF reconstructions of Lockwood, Rouillard \& Finch (2009\nocite{lockwood_rouillard_finch09}), and Svalgaard \& Cliver (2010\nocite{svalgaard_cliver10}).  We find a very good correlation between HMF and polar flux during periods of solar minimum, as well as a good correlation between the square root of sunspot area and HMF during periods of activity.

Finally, through the application of an F-test, we find that the combined explicative power of polar flux and sunspot area is larger than sunspot area alone (with a statistical significance higher than 99\%).  Our results show that the strength of the HMF at Earth is directly related to the dipolar moments (axial and equatorial) of the Sun's magnetic field and those in turn can be quantified by the square root of sunspot area and polar flux -- verifying the theoretical results of Wang \& Sheeley (2003\nocite{wang-sheeley03}) and Wang, Lean \& Sheeley (2005\nocite{wang-lean-sheeley05}).  Considering the causal relationship between the solar surface magnetic field and the evolution of the heliospheric environment, the results of this work represent a step forward towards the practical goal of being able to forecast near-Earth space conditions as our models of the long-term evolution of the surface magnetic field improve; this paves the way for a better understanding of long-term solar variability.

\acknowledgements

\section{Acknowledgements}

We are grateful to Steve Cranmer for the original suggestion which led to this collaboration, Georgios Cintzoglou for his invaluable help with the automatic detection algorithm, the anonymous referee for feedback which helped improved the quality of this paper, Aad Vanballegooijen and Yi-Ming Wang for useful discussions, Mike Lockwood and Leif Svalgaard for sharing their HMF data, and discussing their usage with us, Laura Balmaceda and Sami Solanki for sharing their sunspot database with us, and Jeneen Sommers for her invaluable help in acquiring MDI intensitygram and LOS magnetogram data.  This research is supported by the NASA Living With a Star Jack Eddy Postdoctoral Fellowship Program, administered by the UCAR Visiting Scientist Programs and has made extensive use of SAO/NASA's Astrophysics Data System.   The computations required for this work were performed using the resources of the Harvard-Smithsonian Center for Astrophysics -- we thank Jonathan Sattelberger and Alisdair Davey for much appreciated technical support.  Wilcox Solar Observatory data used in this study was obtained via the web site \href{http://wso.stanford.edu}{http://wso.stanford.edu} at 2011:11:7 courtesy of J.Todd Hoeksema.  The Wilcox Solar Observatory is currently supported by NASA. The MDI instrument is part of SOHO, which is a project of international cooperation between ESA and NASA.  OMNI data were obtained from the Space Physics Data Facility at NASA's Goddard Space flight center \href{http://omniweb.gsfc.nasa.gov/}{http://omniweb.gsfc.nasa.gov/}.  Andr\'es Mu\~noz-Jaramillo is very grateful to David Kieda for his support and sponsorship at the University of Utah. Neil R.\ Sheeley is grateful to Roger K.\ Ulrich (UCLA) for permission to examine the historical collection of Mount Wilson white light images, and to his colleagues J.\ E.\ Boyden and S.\ Padilla for help during the most recent data reduction campaign.  At NRL, financial support was provided by NASA and the Office of Naval Research.  Ed DeLuca was supported by contract SP02H1701R from Lockheed Martin to SAO. Jie Zhang is supported by NSF grant ATM-0748003 and NASA grant NNX07AO72G.

\bibliographystyle{apj}

\newpage

















\end{document}